%  Photon12 copies corrected 1/9/15
%2/8/15 replaced 1.13 ... corrections 2/19 first
%2/20
%copied Dynam4 to 5. 3/30/15 typed 3 pages.

\magnification=\magstep1
\input amstex
\documentstyle{amsppt}
\pageheight{8.5truein}
\topmatter
\title {Dynamics and Scattering of a Massless Particle}
\endtitle
\author   Avy Soffer \endauthor

\address Math Dept. Rutgers Univ. NJ, USA. \endaddress
\abstract

We estimate the probability of a photon to move faster than light,
under the Einstein dynamics which, unlike the wave equation or Maxwell wave dynamics,
has singular dispersion relation at zero momentum. We show that this
probability goes to zero with time, using propagation estimates
suitably multi-scaled to control the contribution of low frequencies.
  We then prove minimal velocity bounds.
\endabstract
\endtopmatter
\document

 \head {Section 0. Introduction}
\endhead
\medskip

Let $F(\lambda > c)$ stand for a smoothed out characteristic
function of the interval $\lambda > c $. We say that a quantum
system, with a configuration coordinate $x\in \Bbb R^3$, obeys a maximal
velocity bound, if for any solution $\psi( x,t)$, localized in
a compact energy interval , there is a positive constant $c$, s.t.
$$
\| F \left( \frac{|x|}{t} >c \right) \psi( x,t) \| _{L^2} \to 0,
$$
as $|t| \to \infty.$

Here we use the notation (Property F)
$$
F(\lambda > c) \equiv \cases 1 & \lambda > c + \delta \\
                             0 & \lambda \leq c-\delta \endcases,
                             \quad \quad  0\leq F, F^\prime \in C^\infty.\tag F
$$

It is well known that such bounds hold for
non-relativistic $N-$particle quantum systems (see \cite{SS, HSS}).
This  bounds are still open for massless particles interacting with
such systems.  See however [Kit, GJY] and the recent works [BFSig,FSig1-2]. The goal of the present paper is to prove such bounds
for a model of a single photon (and some simple generalization) with an effective interaction
 described by a potential $V(x)$.

 New tools will be developed to treat this problem and similar systems.
  The main difficulty here is the singular nature of the symbol of the corresponding hamiltonian,
  given by $|\xi|$ near $\xi=0$.
  This behavior is typical to systems with massless particles (photons of QED, neutrinos of QCD, quasi-particles in solid state systems, e.g. rotons and more).
  The breakdown of the standard  commutator methods was dealt with in several previous works by the introduction of non self-adjoint multipliers(conjugate operators) and/or mollifying the interaction terms near zero frequencies [BFSig,FSig1-2 and cited references].

  The approach developed here, which extends standard positive commutator methods to singular symbols may be of general interest; these methods can be used at the thresholds of the operators in question, where the Mourre theory or propagation estimates fail in general. In particular we derive {\it quantitative Mourre estimates} at thresholds for certain classes of hamiltonians.

  The importance of solving this problem stems from its direct application to various approximate models of QED and other QFT theories with massless particles.
As shown very recently [FSig1-2], one key step to proving velocity bounds for photons of non-relativistic type QED models, is a proper construction of a positive commutator at the one particle massless case, which can then be lifted to the Fock space by the standard second quantization formalism. Furthermore, the rigorous analysis of the photo-electric effect, even at the level of finite particle systems is still open[GZ and cited references].

The method we use in this work is to localize dyadically the energy/frequency as we approach the singularity at zero. On each dyadic interval, we can construct propagation observables and get the desired estimates, since the energy is away from zero frequency (by an amount of order $2^{-n}$, $n$ large). Furthermore, since we are away from zero the derivatives of the symbol are bounded, though we still lose powers of $2^n$.
Then we add up the contributions from all dyadic energy intervals, and use the adapted propagation estimates to each such interval.
A crucial observation is that by having the initial data localized in space, the contributions from each interval shrink to zero as n goes to infinity, due to (an extension of ) Hardy's inequality that controls $E_n(H)<x>^{-1}$ by order $2^{-n}.$
Here $E_n(H)$ is the projection of the operator $H$ on the dyadic interval $[2^{-n},2^{-n+1}].$

  The model we study is given by
 the time-dependent Schr\"odinger equation

$$
\imath \frac {\partial \psi}{\partial t} = H\psi \quad \quad  \psi|_{t=0}
= \psi_0, \tag 0.1a
$$
with the Einstein-Schr\"odinger self-adjoint operator $H$ on $L^2
(\Bbb R^3)$ with Domain $H^1 (\Bbb R^3)$,  given by
$$
H := (-\Delta)^{1/2} + V(x)\equiv |p|+ V(x). \tag 0.1b
$$
This Hamiltonian is a simplified, scalar field version of the
restriction of the photon field in QED, to the one particle sector [Ger,GJY].
We think of the equation (0.1) as a model describing dynamics of a
single photon.
 Moreover, it is used as a laboratory for developing the methods to study propagation with singular dispersion
 at zero energy, needed in tackling the full non-relativistic QED problem.
 Notice that the equation without the interaction term, is
 relativistic invariant. Moreover, by multiplying the equation by
 $-i\partial_t -|p|$, we get the wave equation. However, since the
 initial data for the derived wave equation, is not localized:
 $$
 \dot \psi(0)=|p|\psi(0),
 $$
 we can not use the finite propagation property of the wave equation
 for this model.

 We make the following assumptions:

(i) $V(x) ,$  $  x\cdot \nabla V(x), <x>\Delta V(x)$ are sufficiently regular and
 decay faster than $O(|x|^{-2})$ at infinity;

 (ii) $H$ has no zero energy resonances or zero energy
 eigenvalues.

% $\psi_0 \in L^2 \bigcap L^2 ( \Bbb R^3, |x|^ \alpha d^3x)$, \text{ for some} $\alpha > 0.$

%Our goal is, to prove a decay estimate for: $$ \| F \left( \frac{|x|}{t} >R \right) \psi( x,t) \| _{L^2},$$ for $ R > 1, $ \text{and} $|t| \to \infty.$
The main result of this paper is the following theorem:

\proclaim{Theorem}
 %For $H$ as defined in Eq. (1.1b), in three space dimensions, we  assume moreover, that
 Let $\psi(x,t)$ be the solution of the Schr\"odinger equation (0.1), and assume, furthermore,
 that $H$ satisfies the conditions (i) and (ii). If the initial data
 is localized in  energy in some compact interval, and is such that
 $ <x>^{1+\epsilon} \psi_0$ has a bounded $L^2$ norm, we have the
 following  maximal velocity bound:
 $$
\| F \left( \frac{|x|}{t} >R \right) \psi( x,t) \| _{L^2}=
 o(1)\| \langle x\rangle^{1+\varepsilon} \psi_0 \|_{L^2} \tag
0.2
$$
for $ R > 1 $ \text{and} $|t| \to \infty.$
Similar estimates hold for$|x|/t<\theta<1.$
\endproclaim

 As was mentioned above, the proof of such a propagation estimates for general $N-$body Hamiltonians is known,
%standard, once the Mourre estimate is proved
 see e.g. [Sig-Sof, HSS].
%, by the use of time dependent propagation estimates. These has
%been generalized to some extent,
%to include (mainly) non-relativistic QFT and,  in particular
%non-relativistic QED.
 See also [BFS and cited references]. However, the fact that
the one particle Hamiltonian of a massless photon is singular at zero
energy, prevents a direct extension of the general theory of
propagation estimates to the field, see however [BFSS, FGS,Kit].

The proof is based on the construction of propagation observables,
on each (dyadic) scale of the energy. Suppose that the length of the
momentum $p=- i\nabla, |p|$, is localized in $ [2^{-n-1},2^{-n}],\quad
n\geq0$. Consider then the operator with $F$ having the property (F):
$$
F(A/Rt2^{-n}>1)
$$
defined through the spectral theorem, with $A$ is the self-adjoint
generator of dilation
$$
A=-\imath /2(x\cdot \nabla +\nabla \cdot x), \tag 0.4
$$
and $R>1$.

We start from the basic identity, derived from the Schr\"odinger
equation:
$$
\partial_{t} \langle \psi(t),F\psi(t)\rangle = \langle \psi(t),\{i[H,F]+\partial_{t}
F\}\psi(t)\rangle  \equiv \langle \psi (t), D_H F \psi (t)\rangle.\tag 0.5
$$
Here, $F= F(t)$ stands for any family of self-adjoint operators, for which the
differentiation can be justified, for the chosen $\psi(t)$. To get a
useful propagation estimate, we use the idea of negative propagation
observables,[Sig-Sof]: Suppose $ F$ stands for a family of time
dependent operators, bounded from above (e.g., negative), and such
that its Heisenberg derivative, $D_HF$, defined above, is positive up to
integrable (in time) corrections. We will then obtain, upon
integration over time of (0.5), the following propagation estimate:
$$
\int_{1}^{T} \|B\psi(t)\|^2 dt\leq C\|\psi(0)\|^2 .\tag 0.6
$$
Here, we use that the Heisenberg derivative, $D_{H}F$, is given by a
positive operator, plus integrable corrections:
$$
D_{H}F=i[H,F]+\partial_t F= B^*B + R(t), \int^\infty_1 \langle \psi(t), R(t)
\psi (t) \rangle  dt <C \|\psi (0)\|^2. \tag 0.7
$$
In general, we need to use phase space operators which, while not
pseudo-differential, will have suitable phase space support. If we
choose for $F$, the function of $A$, defined above as our
propagation observable, we need to estimate the commutator of $H$
with this $F$, suitably localized in some energy shell. One can then
use the following basic commutator expansion Lemma:

Let $H$ and $A$ be self-adjoint operators on a Hilbert space $\Cal H$
and suppose that $H$ is bounded.  To say that the commutator $i[H, A]$
is bounded means that the quadratic form
$$\imath [(H\psi, A\psi) - (A\psi, H\psi)]
$$
on $D(A)$ is bounded and thus defines a bounded, symmetric operator called $i[H,
A]$.  In the same sense we assume that the higher commutators
$$
ad^{(k)}_A (H) = [ad^{(k-1)}_A (H), A ]
$$
are bounded for $k=2\dots n$.  Let $f$ be a real $C^\infty$-function
on $R$.  Then, under a further condition given below, the commutator
$[H, f(A)]$ has an expansion:
$$
[H, f(A)]= \sum^{n-1}_{k=1} \frac{1}{k!} f^{(k)} (A) ad^{(K)}_A (H) +
R_n \tag *
$$
with a remainder estimate
$$
\|R_n\| \leq c_n \| ad^{(n)}_A (H)\|\sum^{n+2}_{k=0}\int dx (1+
|x|)^{k-n-1} |f^{(k)}(x)|.
$$
The further condition on $f$ is that the integrals above exist.  The
number $c_n$ is a numerical constant depending on $n$ but not on $f,
A$ or $H$.  In particular, the expansion holds if
$$f^{(k)}(x) = O (|x|^{n-\varepsilon -k}) \quad (x\to \pm \infty)
$$
for $k= 1 \dots n + 2,$ i.e. if the function$f(x)$ grows not faster than
$|x|^{n-\varepsilon}$, with corresponding slower growth of the
successive derivatives.  In that case (*) is defined in form sense on
the domain of $f^{(1)} (A)$.

Therefore,
$$
\imath[H, F(A)]= F^{\prime}(A) \imath[H,A]+ R([[H,A],A]),\tag 0.8
$$
where the remainder is given by an explicit multiple integral,
involving the group generated by $A$, and, the double commutator
above, of $H$ with $A$. By the above expansion, we obtain, through
symmetrization of the first term on the RHS of (0.8), that the above propagation observable has
Heisenberg derivative that is smaller than:
$$
-\frac{1}{t}G_n^{2}(A/t)+R(t,n),  \quad G^2_n (A, /t) \equiv \lambda
 F^\prime (\lambda) |_{\lambda= A/R t 2^{-n}}.\tag 0.9
$$
So, $ G_n$ stands for a bump function of $\frac{A}{Rt2^{-n}}$
around $1$. We will get a useful estimate, if we can control the
remainder term, $R(t,n)$, by an integrable function of $t$, which is
also well behaved as $n$ tends to infinity. The leading term in the
expression (0.9), comes from the time derivative of $F$, of the
Heisenberg derivative. The first term in the Heisenberg derivative,
the commutator, has an expansion beginning with:
$$
F^{\prime}_n (A/R 2^{-n} t)\imath [H, A]\frac{2^n}{Rt}= F^\prime_n (A/R 2^{-n}t) \imath [|p| -2
x \cdot \nabla V ] \frac{2^n}{Rt}, \tag 0.10
$$
which can be shown to be much smaller than the leading term, by
using the energy localization around $2^{-n}$. To see that, we use
that the commutator $\imath [H,A]$ is bounded by $c|p|$, for some finite
constant $c$, by the use of the uncertainty principle:
$$
\align
&| (\psi, i[H, A] \psi) |\\
& =|\left(\psi, ( |p| -2 x \cdot \nabla V)\psi\right) |\\
& \leq (\psi, |p| \psi)\\
& + c(\psi, <x>^{-2-\varepsilon} \psi)\\
&\leq c(\psi, |p| \psi).\endalign
$$
and our decay
and regularity assumptions on $V$. Then, we need to show that energy
localization implies a similar momentum localization. This can not
be done using standard localization arguments, since the derivative
of the localization function grows like $2^n$! To this end, a
completely different argument is used: One proves that under generic
spectral assumptions on $H$, $HP_c(H)$ dominates a constant times
$P_c(H)|p|P_c(H)$. $P_c(H)$ stands for the spectral projection of
$H$ on its
 continuous spectral part.

 It then follows that:
$$
E_n(H)|p|E_n(H) \leq c2^{-n}E_n(H)^2.\tag 0.11
$$
Finally, we need to show that the remainder $ R(t,n)$, higher
commutator terms in the expansion of the Heisenberg derivative, are
integrable in $t$. This is the most involved step; without energy
localization the formal expression for the remainder is given by a
divergent integral. The way to estimate this last remainder term, is
to use the fact that the group generated by dilations moves the
support of the energy localization functions. Consequently, the
integrals over the group actions are limited to finite domains. The
remainder term comes from the following expansion:
$$
\align E_n i [ |p|, \Phi_n ] E_n &= E_n i [ |p|^{1/2} , \Phi_n ]
|p|^{1/2} E_n
+ E_n |p|^{1/2} i [ |p|^{1/2} , \Phi_n ] E_n\\
&= E_n |p|^{1/2} F_n^{\prime}(A/t) |p|^{1/2} E_n \frac{1}{Rt} +2\Re
E_n |p|^{1/2} R_2 (A/t, n) E_n, \tag 0.12
\endalign
$$
with
 $$ \Phi_n \equiv E_{n}F(\frac{A}{tR2^{-n}}>1)E_{n}$$
 and with $R_2(A/t, n)$ given by
$$
R_2(A/t, n) = \frac{1}{2} \int d\lambda \hat F_n(\lambda) e^{i\lambda
A/Rt} \int_0^\lambda ds \int_0^s du \,   e^{-iu A/Rt }  \,
\frac{1}{2} |p|^{1/2}\quad e^{iuA/Rt} (Rt)^{-2}
$$
$$ = \frac{1}{4}
\int d\lambda \hat F_n (\lambda) e^{i\lambda A/Rt} \int_0^\lambda ds
\int_0^s du \quad e^{-u /Rt} |p|^{1/2} \quad (Rt)^{-2}. \tag 0.13
$$
If we try to bound the expression for $R_2$ by taking the norm of
the first term on the RHS of (0.13), as it was done in past works, we lose a
factor of $2^{-n/2},$ coming from localizing the $|p|^{1/2}$ factor.
On the other hand, we can not directly estimate the last expression
on the RHS of equation (0.13), since the integrand grows
exponentially in $\lambda$, while $\hat F_n$ decays slower than exponential,
being the Fourier transform of a compactly supported function. To
this end, we use the fact that the dilation group, generated by $A$,
changes the support of functions of $|p|$, or $H$:
$$
E_{I_n}(|p|)e^{i\lambda A}E_{I_n}(|p|)=E_{I_n}(|p|)
E_{I_n}(|e^{-\lambda} p|) e^{i\lambda A}  = 0, \tag 0.14
$$
for $| \lambda |\ge \ln2.$ Then, we use the mutual domination of $|p|,H$:
$$
P_c(H)H\le cP_c(H)|p|P_c(H)\le dP_c(H)H, \tag 0.15
$$
for some positive constants $c,d$.

 Combining equations (0.14),
(0.15), we can then show that the integration on $\lambda$ is
limited to a compact domain in $\lambda$, in equation (0.13).
 Collecting all of
the above, (0.5 - 0.15), we get estimates of the form:
$$
E_n(H)\frac{1}{t}G_n^2(A/t)E_n(H) \in L^1(dt).\tag 0.16
$$
This estimate is then jacked up by the use of the propagation
observable $ \frac{A}{t}F_n(A/t),$ to obtain,
$$
E_n(H)F_n(A/t)E_n(H)\frac{1}{t} \in L^1(dt).\tag 0.17
$$
In the next step of the proof, we estimate using the above, the
following operator:
$$
E_n(H) F(\frac{|x|}{t}>R)E_n(H).
$$
We write:
$F(\frac{|x|}{t}>R)=F(\frac{|x|}{t}>R)F_n(A/t)+F(\frac{|x|}{t}>R)\bar
F_n(A/t).$
$$
F_n+\bar F_n=1.
$$
The first term of the above decomposition, goes to zero, as time
goes to infinity, by the above propagation estimates, on $F_n(A/t).$

So, we need to show that the $\bar F_n$ term also goes to zero.

This is formally true , since it consists of a product of two
operators, with disjoint classical phase-space support, on the
energy shell $2^{-n}.$

 Again, the proof of this property
necessitates the use of new phase space localization arguments. In
the final step of the proof, we sum over all $n$, and in the process
we lose some powers of $2^{-n}$. These are compensated by requiring
the initial data to be localized in $x$, and by using that(up to 2)
negative powers of $|p|$, are bounded, up to a constant, by positive
powers of $|x|$.

 There are new difficulties in completing this argument, compared
with the usual case, without dyadic energy localization.

 First, we
need to minimize the number of powers of $2^n$, coming from
expanding the function $F_n$.
 Then, we need to trade positive
powers of the momentum( derivative operator) $p =-i \nabla$,  for powers of
$2^{-n}$.

Finally, to control the remainder term in the Commutator Expansion
Lemma, the $Q_2$ term, (or $R_2$ term), we need to commute the
derivative through the dilation group, which produces exponentially
large factors.

 The way out of these problems involves the following
arguments.
To limit the integrations in the remainder term $R_2$, we notice that
the dilation group moves the dyadic energy interval, away from its
original support.
 Hence, for large enough value of the group parameter, $\lambda$,
the fact that our propagation observable is localized on the dyadic
interval, from both sides, gives an extra decay, that cancels the
exponential growth factor. This is shown in detail in the subsection
"The term $R_2$".

 To get the $2^{-n}$ factor from the momentum $p$, we prove some
propositions about the properties of the operator $H$, which might
be of independent interest. (see Proposition (2.7))
 Specifically, we show, that in three (and higher)
dimensions, if there are no zero energy resonances and eigenvalues,
then $H$ and $|p|$ dominate each other, up to a multiplicative
constant, on the continuous spectral subspace of $H$.

 These estimates are the key to getting the
right minimal powers of $2^n$, from the various propagation estimates
and phase space localizations.
\smallskip

\medskip

\head {\bf ACKNOWLEDGEMENTS} \endhead

I wish to thank I.M. Sigal for useful discussions. Part of this work
was done while the author visited the IHES, France. A. Soffer was
partially supported by NSF grant number DMS-0903651 and DMS-1201394.

\head {Section 1. Propagation Estimates}
\endhead
\medskip
Our goal in this section is to prove the following key propagation
estimate, as sketched in the introduction:
$$
\int_{1}^{T} \|F_n(A/t)E_n\psi(t)\|^2 \frac{dt}{t}\leq
C\|E_n\psi(0)\|^2 .\tag 1.1a
$$
and with

$$
F_n(A/t)\equiv F(\frac{A}{Rt2^{-n}}>1). \tag 1.1b
$$
$E_n$ stands for the operator $E_{I_n}(H)$.
 We use propagation
estimates, with the propagation observables
$$
    \Phi _n \equiv E_ {I_n} (H) F\left( \frac{A}{tR {2^{-n}} } > 1 \right) E_ {I_n} (H), \tag
    1.2a
$$
$$
\Phi^{(1)} _n \equiv E_ {I_n} (H) \frac{A}{t}F\left( \frac{A}{tR {2^{-n}}
}
> 1 \right) E_ {I_n} (H), \tag 1.2b
$$
where, $    n = 0,1, \cdot \cdot
\cdot \infty , $ $I_n$ stands for the interval $[2^{-n-1} ,
2^{-n}]$, $E_ {I_n} $  is  the characteristic function of  $I_n $
and
 $A$ is the dilation generator,
$$
    A \equiv \frac{1}{2} (x \cdot p + p \cdot x). \tag 1.3
$$
On the support of $F$, at least classically, $|x|\ge \delta t$, since $p$ can not be unbounded on the support of $E_n(H)$.
Consequently, we can drop ,to leading order the contribution of $V(x)$ to $H$. So, in fact, to leading order, $E_n(H)\thickapprox E_n(|p|).$
Therefore at the classical level, $ \frac{|x|}{t} \geq R $ on the support of $\Phi_n, $
 and $  R\leq \frac{|x|}{t} \leq 2R$ on the support of $E_nF^{\prime}E_n . $

Furthermore,
$$
    |\partial^j_\lambda F \left( \frac{\lambda}{Rt2^{-n}} > 1 \right)| \sim \left( \frac{2^n}{Rt} \right) ^j . \tag 1.4
$$
%where the derivatives are taken w.r.t.  the variable $\lambda$.

The main propagation estimate is based on showing that:
$$
\align
& E_{I_n} i \left[ |p| + V, F \left( \frac {A}{Rt2^{-n}}
> 1 \right) \right] E_{I_n}\\
&=E_n \left \{ i |p|^{1/2} \left[ |p|^{1/2} , F \right] + i\left[
|p|^{1/2}, F \right] |p|^{1/2} + i[V,F] \right \}E_n \\ &= E_n \left
\{ |p|^{1/2} \tilde{F}^2 |p|^{1/2} + |p|^{1/2} \frac
{2^{2n}}{R^2t^2} Q_2(A/t) |p|^{1/2} \right \}E_n+O(t^{-2})  ,  \tag
1.5a
\endalign
$$
where $\tilde{F}^2 = \frac{2^n}{Rt} F^{\prime}$ \quad and with the bounds:
$E_nQ_2(A/t)E_n =O(1)$ for $t \ge 2^n/R,$ and of order
$2^{-n},$ for $t \le 2^n/R.$
 Then, we show that
$$    E_n |p|^{1/2} \tilde{F}^2 | p|^{1/2} E_n \lesssim (c/Rt) E_n
G_n^2 E_n,\tag 1.5b
$$ with
$$
G_n^2=F(\frac{A}{Rt2^{-n}}\sim 1). $$

Letting $ F_n(\lambda)=F(2^{n}\lambda \ge 1)$ and using that
$-\frac{A}{t}F_n^{\prime} \sim -R2^{-n}F_n^{\prime}, R\gg1 $, we
obtain, using also
$$
\align
\int _1^T\partial_t < \psi_t, E_{I_n}(H)F_n\left(\frac{A}{Rt}\right)
&E_{I_n}(H)\psi_t > dt = \int_1^T  <\psi_t, \imath [H,\Phi_n]\psi_t >dt
 \\
&+
2^{n}R^{-1}\int_1^T<\psi_t,E_{I_n}(H)(\frac{-A}{t^2})F_n^{\prime}E_{I_n}(H)\psi_t>dt.\tag
1.6
\endalign
$$
 that bounds (1.5),(1.6) imply:

\proclaim{Theorem 1.1}
 Under the previous assumptions on $H$, and
assume that (1.5) holds, we have the following propagation
estimates:
$$
\int_1^T \|F_n^{\prime}(A/Rt)E_n\psi(t)\|^2 \frac{dt}{t} \le
c(R)\|E_n\psi(0)\|^2\tag 1.7a
$$

$$
\align
 \int_1^T \|G_n (A/Rt)E_n\psi(t)\|^2& \frac{dt}{t} +<\psi(T),E_n
\frac{A}{T}F_n(A/RT)E_n\psi(T)> \\ \le c(R)
 \|
<A>^{1/2}E_n \psi(0)\|^2 \tag1.7b
\endalign
$$

\endproclaim

\demo{Proof}

 To prove (1.7a), we note that it is the same as (0.6) with
$$
B=\tilde{F}_n (A/t) E_n(H).
$$
As explained in the introduction, (0.6) follows upon integration over
time, from $0$ to $T$ of the identity (0.5) and application of the
fundamental theorem of calculus, provided that we prove the following
commutator estimate:
$$
\aligned
D_H\Phi_n & = D_H E_n(H)F_n\left(\frac{A}{tR} >1\right)E_n (H) \\
          & =-\frac{\tilde{c}(R)}{t} E_n (H) \tilde{F}^2_n (A/t) E_n (H) +
O\left(L^1(dt)\right).\endaligned \tag 1.8
$$
With this choice of $\Phi_n$, the integral over time of (0.5) takes
the form (1.6).

Using the definition that
$$
G^2_n (A/t) \equiv 2^n \frac{A}{Rt} F^\prime_n
(A/t),\tag 1.9
$$
the second term on the RHS of (1.6) is seen to be equal to
$$
- \int^T_1 \langle \psi (t) ), E_n (H) \frac{1}{t} G^2_n (A/t) E_n
(H) \psi(t) \rangle dt \tag 1.10
$$
which corresponds to the first term on the RHS of (1.8).  Therefore,
the main estimate is to show that the correction terms are in
$L^1(dt)$, uniformly in $2^n$.  This is achieved by proving Reverse
Mourre Estimate [Sig-Sof]:

$$
\| E_n i[H, A] E_n \| \lesssim 2^{-n}.\tag 1.11
$$
We also need to control from above the second order correction term in
the expansion of the commutator $i[H, F_n (A/t)]$ by the commutator
expansion Lemma.

These are proved in Proposition A, B, together with the general
inequalities of Section 2.  To proceed, note that the commutator in
the correction term in (1.6) - the first term, is expanded in (1.5a)
as a sum of three terms: $\tilde{F}^2$ term, $Q_2$ term and
$O(t^{-2})$.  The first term, the $\tilde{F}^2$ term, is dominated
(for $R> c>1)$ by a small constant times the leading term in 1.8,
since
$$
\tilde{F}^2 = \frac{2^n}{Rt} F^\prime_n (A/t) = \frac{1}{R} G^2_n (A/t)
$$
and $E_n (H) |p|^{1/2} = O(2^{-n/2})$, (Section 2, by Prop 2.7).
Next, we consider the remainder terms $Q_2$ and $O(t^{-2})$.  These
terms come from the expansion of the commutator
$$
E_n i[H, F_n (A/t)] E_n = E_n i [V, F_n (A/t)] E_n + E_n i
[|p|, F_n(A/t)]E_n. \tag 1.12
$$
\enddemo

%\head

We proceed by proving some Preliminary Estimates.
%\endhead
\medskip

%1/13
\medskip

\proclaim {Lemma 1.2} For $W(x)$, smooth and decaying fast enough,

$$
B_n \equiv \| F_n (A/t) W(x) E_n (H) \|\lesssim
\frac{1}{Rt}\tag 1.13
$$

\endproclaim

\demo{Proof of Lemma}

%replaced with red page 21??    already replaced    ???
%??   not sure (*2) where it goes after. moved pg13

$$
\aligned
B_n & = \frac{1}{2} \| F_n (A/t) A^{-1} (x\cdot p + p\cdot x) W(x) E_n (H) \|\\
    & = O(2^n/Rt) \left\{ \|(x W^\prime  + W + xW^\prime) E_n (H) \|\right. \\
    &\left.  + 2\| x W (x) p  E_n (H) \|+ \| W(x) E_n (H) \|\right\}\\
    & = O(2^n/Rt) \left\{ \|\langle x \rangle^{-1} E_n (H) \| \, \|
    |xW| + |x^2 W^\prime \|_{L^\infty}\right\} \\
&\leq \left(\frac{1}{Rt}\right).
\endaligned  \tag 1.14
$$

\enddemo

We used that $A^mF_n (A/t) = O \left([2^{-n} / Rt]^m\right) F_n
(A/t)$,and
$\|pE_n(H)\|+\|<x>^{-1}E_n(H)\|=O(2^{-n}.$  which proves the Lemma.

$\blacksquare$

Therefore, using $F_n (A/t) A^{-2} = \tilde{F}_n (A/t) 2^{2n}/t^2$ we
get from the above computation that
$$
\aligned
\| F_n (A/t) W (x) E_n (H)& \|\lesssim (2^{2n}/ R^2 t^2)\times\\
 &\|\langle x \rangle^{-1} E_n (H) \|  \|\langle
x \rangle^2 |W| + \langle
x \rangle^3(| W^\prime| + | W^{\prime\prime}|)  \|_{L^\infty}\\
&\leq c (2^n/R^2t^2)
.\endaligned
\tag 1.15
$$

  From this we immediately conclude that
$$
\int^\infty_{\varepsilon 2^n}  dt \langle \psi(t), E_n (H) i [V, F_n (A/t)] E_n
(H) \psi (t)\rangle \leq \frac{1}{\varepsilon R^2} \| | V | \| \quad
\|E_n(H)\psi \|^2.  \tag 1.16
$$

The shortime estimate, $t\leq 2^n /R$, follows from the bound:

$$
\aligned
 & \big|\langle E_n \psi(t), (H F_n - F_n H) E_n (H) \psi (t) \rangle
\big|\\
 & \leq 2 \|H E_n \psi (t) \| \quad \| F_n \| \, \| E_n \psi (t) \| \\
 & \leq 2\cdot  2^{-n} \|E_n (H) \psi (t) \|^2.\endaligned \tag 1.17
$$
$$
\aligned
\Rightarrow \big| &\int^{2^n/R}_0 \langle E_n (H) \psi(t), i[H, F_n
 (A/t)] E_n (H) \psi (t)\rangle dt \big|\\
 & \leq C\|  E_n (H) \psi (t) \|^2/R.  \endaligned \tag 1.18
$$

Next, we consider the second term on the RHS of (1.12):
$$
E_n i [ |p|, F_n(A/t)] E_n \tag 1.19
$$
The estimate (1.18) applies to the left side of eq. (1.12), so, we
only need to estimate the above commutator for $Rt > 2^n$.

The above commutator can be written as (1.5a), after using that
$$
i[|p|, F_n] = i|p|^{1/2} [|p|^{1/2}, F_n] + i[|p|^{1/2}, F_n]
|p|^{1/2}\tag 1.20
$$
and then expanding the commutators, using the commutator expansion
Lemma:
$$
\aligned
i[|p|^{1/2},  F_n] &= F^\prime_n i\left[|p|^{1/2}, \frac{A}{Rt
2^{-n}}\right]+R_2\\
   &=(Rt)^{-1} 2^n F^\prime_n(A/t)(+\frac{1}{2} |p|^{1/2}) + R_2.
\endaligned\tag 1.21
$$

Since $F^\prime \geq 0$, we can write
$$
\tilde{F}^2 \equiv \frac{2^n}{Rt} F^\prime,\tag 1.22
$$

Furthermore, by taking the adjoint of eq. (1.21) we get
$$
i[|p|^{1/2}, F_n]= \frac{1}{2} |p|^{1/2} \tilde{F}^2 + \bar{R}_2.
$$
(1.20 - 1.22) implies eq. (1.5a).  For $R$ large enough, the first
   term on the RHS of (1.5a), while positive, is dominated by the
   leading term (1.10).  So it remains to control the integral over
   time of the $R_2$ terms, only for $t> 2^n/R$.  For $t\leq 2^n/R$ we
   bound the whole commutator $i[H, F_n]$, see eq. 1.18.

This is shown in propositions A, B.  We now consider the proof of
(1.7b).  To this end, we employ a similar proof as for (1.7a), but with
a new propagation observable
$$
\Phi^{(1)}_n \equiv \frac{A}{t} F_n
(A/t). \tag 1.23
$$
The left hand side is now an {\bf unbounded} operator, but nonnegative,
since $F_n(x)$ is supported in $x\geq 2^{-n} R$.  Hence, the
argument leading to (1.18), for $t\leq 2^n/R$ does not apply.  The
proof in this case is done in the next section, proposition D.  The
RHS consists, as in eq. (1.6) from a commutator term and a derivative
w.r.t time.  The derivative w.r.t. time gives
$$
-\frac{A}{t^2} F_n (A/t) - \frac{A}{t^2} \frac{2^n}{R} F^\prime_n (A/t) \leq
 0.\tag 1.24
$$

The expression in (1.24) is easily seen to be less than

$-\frac{A}{t^2}
F_n (A/t) - \frac{1}{t} \tilde{F}^2(A/t) \leq -  \frac{R}{t} F_n$
with
$$
\tilde{F}^2(x) = F\left(\frac{2^n x}{R} = 1\right), \tag 1.25
$$
that is, a $C^\infty_0$ bump function around $R2^{-n}$.

The commutator term will be shown to be dominated by
$$
\frac{2^{-n}}{t} F_n (A/t) + \frac{1}{Rt} \tilde{F}^2 (A/t) \tag 1.26
$$
in proposition C.

Hence (1.26) is dominated by (1.25) for $R>1$.  The higher order terms
$R_2$, are controlled in proposition C.  Upon integration overtime,
the bound (1.7b) follows.

$\blacksquare$

We now proceed to prove the statements that were assumed above.  This
is the content of the next 4 propositions A-D.

We begin with estimating the $E_n (H) R_2 (H, F_n(A)) E_n (H)$ for $t >
R^{-1} 2^n$.

%endproof ??

%\head {\bf The $R_2(A/t)$ term.}\endhead

%insert ????

%change to 1.27 ??
%Makes no sense.

\medskip
\proclaim{Proposition A (The $R_2(H, F_n)$ term) }

For $\Phi_n$ as in (1.2a), we have:($Q_2$ is defined via equation (1.5a))
$$
\int_{2^n/R}^T <E_n|p|^{1/2}Q_2|p|^{1/2}E_n> dt \le
 \frac{c}{R}\|E_n\psi(0)\|^2. \tag 1.27
$$

\endproclaim
\demo{Proof}

Direct application of the commutator expansion lemma gives:

$$
 i \left[ |p|^{1/2}, F_n(A/t) \right]
 = \int d\lambda \hat F_n (\lambda) e^{i\lambda A/Rt} \frac{1}{Rt} \int_0^\lambda ds e^{-isA/Rt}
  i\left[|p|^{1/2}, A  \right] e^{isA/Rt} $$$$
= \frac{1}{2} F_n^{'}(A/t) \frac{1}{Rt} |p|^{1/2} + \tilde R_2 (A/t)
|p|^{1/2} = \frac{1}{2} F^{\prime}_n(A/t) \frac{1}{Rt} |p|^{1/2} +
R_2(A/t), \tag 1.28
$$
where we used that
$$
    -i \left[ A, |p|^{1/2} \right] = p \cdot \nabla_p |p|^{1/2} = \frac{1}{2} |p|^{1/2}. \tag
    1.29
$$

$$
R_2(A/t) = -\frac{1}{2} \int d\lambda \hat F_n(\lambda) e^{i\lambda
A/Rt} \int_0^\lambda ds \int_0^s du \quad  e^{-iu A/Rt} \frac{1}{2}
|p|^{1/2}\quad e^{iuA/Rt} (Rt)^{-2}$$$$ = \frac{1}{4} \int d\lambda
\hat F_n (\lambda) e^{i\lambda A/Rt} \int_0^\lambda ds \int_0^s du
\quad e^{+u /Rt} |p|^{1/2}(Rt)^{-2}. \tag 1.30
$$
We used $e^{i\alpha A}|p| e^{-i\alpha A} = e^{-\alpha}
|p|.$
In general, this integral blows up at infinity, due to the fact that $e^{u/Rt}$ grows exponentially fast, while
  $\hat{F}_n(\lambda)$  decays faster than any polynomial, but not exponentially,
  since $\hat {F}_n(\lambda)$ is the Fourier transform of a compactly
supported function. Applying energy localization by $E_n \equiv E_n
(H)$

$$
\align
E_n i [ |p|, \Phi_n ] E_n &= E_n i [ |p|^{1/2} , \Phi_n ] |p|^{1/2} E_n
+ E_n |p|^{1/2} i [ |p|^{1/2} , \Phi_n ] E_n\\
&= E_n |p|^{1/2} F_n^{\prime}(A/t) |p|^{1/2} E_n \frac{1}{Rt} + E_n
|p|^{1/2}  2  Re R_2 (A/t) E_n\\
&\equiv E_n \frac{|p|^{1/2}}
{Rt} F^\prime_n (A/t) |p|^{1/2} E_n\\
& + E_n |p|^{1/2} Q_2 |p|^{1/2} E_n  \tag 1.31
\endalign
$$
with $R_2(A/t)$ given by (1.30).  Let $\tilde{E}_n(\cdot) E_n (\cdot)
= E_n (\cdot)$.

$$
E_n (H) = E_n (H) - E_n (|p|) + E_n (|p|) \equiv \delta E_n + E_n
(|p|).
\tag 1.32
$$
Hence, the $R_2(A/t)$ term becomes
$$
\aligned
E_n(H) |p|^{1/2} R_2 (A/t) E_n(H) &= |p|^{1/2} E_n(|p|) R_2(A/t) E_n
(|p|)\\
 &+ \delta E_n |p|^{1/2} R_2(A/t) E_n(|p|)\\
 &|p|^{1/2} E_n (|p|) R_2 (A/t) \delta E_n\\
 &+\delta E_n |p|^{1/2} R_2 (A/t) \delta E_n\\
 &\equiv J_1 + J_2 + J_3 + J_4.
\endaligned\tag 1.33
$$
%previously 2-8-15

$$
\bar {E}_n \equiv 1 - \tilde {E}_n .
$$
In our case,  $ 1 - \tilde {E}_n = \tilde {E} (|p| \leq 1 ) - \tilde {E}_n$.
$\tilde {E}_n$  stands for smoothed $E_n$ function of $|p|$, and where
we used proposition (2.4d) to get that $\bar{E}_n (|p|) E_n (H) =
\bar{E}_n (|p|) O (n2^{-n}) E_n (H)$.

The estimate of J4 follows from that of J2, J3, as it is higher order
by a factor of $n2^{-n}$.

$$
\align \int_1^T J_1 dt&=\int_1^T  |p|^{1/2} {E}_n (|p|) R_2(A/t)
{E}_n (|p|)   dt \\
&=\int_1^T \frac{dt}{R^2t^2}|p|^{1/2}  {E}_n (|p|) Q_2
(A/t) {E}_n (|p|)|p|^{1/2}  \\
&= \int_1^{c2^n} \frac{dt}{R^2t^2}  |p|^{1/2} {E}_n (|p|) Q_2
(A/t) {E}_n (|p|)|p|^{1/2}  \\
&+ \int _{c2^n}^T \frac {dt}{R^2t^2}   |p|^{1/2} {E}_n
(|p|)Q_2(A/t){E}_n (|p|)|p|^{1/2} .\tag 1.34
\endalign
$$
If $T \leq c2^n$ , then the second term on the r.h.s of ( 1.34) is
zero.
\smallskip
We now estimate the the LHS of equation (1.34).
 Using
proposition (2.4c), it follows that the $\lambda$ integration (and
therefore the other integrations) in the expression for $R_2(A/t)$,
eq. (1.30), is limited to
 $$
    |\lambda| \leq Rt \ln2. \tag 1.35
$$
Hence,
$$
J1 =\frac{c}{(Rt)^2} |p|^{1/2} {E}_n(|p|)
\int_{|\lambda| \leq Rt \ln2} d\lambda\hat{F}_n (\lambda)
e^{i\lambda A/Rt}
 $$
$$
\times\int_0^\lambda ds \int_0^s du\,  e^{-iu A/Rt} |p|^{1/2} e^{+iu
A/Rt} {E}_n(|p|). \tag 1.36
$$

$$
\align
J1= \frac{c}{(Rt)^2}
&|p|^{1/2} {E}_n (|p|) \int_{|\lambda| \leq Rt \ln2} \hat {F}_n (\lambda) e^{i\lambda A/Rt}
 \int_0^\lambda ds \int_0^s du \,  e^{+u/Rt}
 |p|^{1/2}{E}_n (|p|) \\
&=\frac{O(1)}{(Rt)^2} E_n(|p|)  2^{-n/2}( \int_{|\lambda| \leq Rt \ln2} |\lambda^2 \hat {F}_n (\lambda)|
 d\lambda) \quad 2^{-n/2} E_n (|p|) \\
&= \frac{O(1)}{(Rt)^2}  E_n (|p|) 2^{-n} (Rt)^2 E_n (|p|) \text{
since }\int |\hat{F} (\lambda) |d \lambda = O(1) .
\endalign
$$
Hence,
$$
    \int_1^{2^n/R}  J1 \quad dt \leq O(\frac{1}{R}).
$$
If  \quad $Rt > 2^n$, we use instead, that
$$
    \int |\lambda^2 \hat {F}_n (\lambda)| \quad d\lambda \leq 2^{2n},
$$
so that,
$$
    J1 = \frac{O(1)}{(Rt)^2} E_n 2^n E_n , \quad \text {and then},
$$
$$
    \int_{2^n /R} ^T  J1 dt = O(\frac {1}{R}).\tag 1.37
$$

\medskip

next, we estimate the the terms $J2-J4$ defined in equation (1.33).
Consider the region \quad $Rt > 2^n $.
 The integrand to estimate can be
written as
$$
(Rt)^{-2 } \hat {F}_n(\lambda) \tilde{J}(\lambda,s,u),\tag 1.38
$$

$$  \tilde {J}(\lambda,s,u) \equiv |p|^{1/2} {E}_n|(p|) e^{i\lambda A/Rt} |p|_u ^{1/2}
     \delta {E}_n,\tag 1.39
$$

$$
\left ( |p|_s \equiv e^{-iAs/Rt}|p| e^{iAs/Rt} \right),\tag 1.40
$$
and adjoint of such term.

First, we decompose the region of integration $\lambda$ to:

$$
    \frac {|\lambda|}{Rt} > m\
    \text {and }  \frac {|\lambda|}{Rt} \leq m,\
    m > \ln2.
$$

For $\frac{|\lambda|}{Rt} > m$, we consider the case $\lambda > 0,
\lambda < 0$ separately.

For $\lambda < 0$, we have $(e^{-iu A/ Rt} |p|^{1/2} e^{+iu A/ Rt} =
e^{u/2Rt})$ since $\lambda < 0$ implies $\lambda \leq u, s \leq 0$,
$$
\aligned
& \|\tilde{J}_2 (\lambda, s, u) \| = \| |p|^{1/2} E_n (|p|) e^{i\lambda
A/Rt} e^{u/2Rt} |p|^{1/2}\delta E_n \|=\\
& =\|O (2^{-n/2}) \| |p|^{1/2} (E_n (|p|) - E_n (H))\| = O (2^{-n}),\endaligned \tag
1.41
$$
and where we used that $e^{i\lambda A/Rt}$ is bounded on $L^2$, and
$e^{u/2Rt} \leq 1,$ for $u\leq 0$.

So, the integral of this part $(\lambda < 0)$ in $J_1$ is bounded by
$$
\aligned
&\int^T_{2^n/R} \frac{dt}{(Rt)^2} \int_{\frac{\lambda}{Rt} \leq -m}
|\hat{F}_n (\lambda) |\lambda^2 O(2^{-n}) d \lambda\\
& \leq 2^{-n} R^{-1}
 O(2^{-n}) \cdot 2^{2n} O (|m|^{-k}) = O (|m|^{-k})\endaligned \tag 1.42
$$
where we use that
$$
\int^T_{2^n /R} \frac {dt}{R^2 t^2} \leq c 2^{-n} R^{-1}, \tag 1.43
$$

$$
\int_{|\lambda | \geq
(2^{+n}/R) m} |\hat{F}_n (\lambda)|\lambda^2 d \lambda  \leq cR /m.
\tag 1.44
$$
Next, we consider $\lambda \geq 0, \lambda /Rt > m$.

In this case, $0 \leq u \leq \lambda$, and therefore $e^{u/2Rt}$ is
large.

So, in this case we commute $e^{i\lambda A/R t}$ to the right
$$
\aligned
\|\tilde{J}_2 (\lambda, s,  u) \| &= \| |p|^{1/2} E_n (|p|)
e^{(u-\lambda)/2Rt}|p|^{1/2} e^{i\lambda A/Rt} \delta E_n\| =\\
 &= O(2^{-n}).
\endaligned
\tag 1.45
$$
The estimate of this part of $J_2$ is therefore the same as the bound
(1.41).

The estimate of $\hat{J}_3$ is identical.

Next, we estimate $J_4$:
$$
\aligned
J_4 &= \delta E_n |p|^{1/2} R_2(A/t) \delta E_n \\
    &= \frac{c}{(Rt)^2} \delta E_n |p|^{1/2} \int_{|\lambda| \geq mRt}
    d \lambda \hat{F}_n (\lambda) e^{i\lambda A/Rt} \int^\lambda_0 ds
    \int^s_0 du |p|^{1/2} e^{i u A/Rt} \delta E_n.
\endaligned
\tag 1.46
$$
The estimate on $J_4$ follows from a sharper bound of $\delta E_n$,
given in the following Lemma:

\proclaim{Lemma 1.3}

Let $E_n(\lambda)$ be as above, a smooth characteristic function of
$I_n$.
$$
\text{ supp } E_n (\lambda) = [2^{-n-1} (1-\delta), 2^{-n}
(1+\delta)].
$$
Then
$$
\|\tilde{E}_n \delta E_n \| = \| \tilde{E}_n(H) (E_n(H) - E_n
(|p|))\leq C2^{-n} |\|V\||_2.\tag 1.47
$$
$$
\| |p|^{\frac{1}{2} + \varepsilon}(\delta E_n) \tilde{E}_n(H)
\|\leq C_\varepsilon 2^{-n-\beta n}
 |\|V\||_{\frac{3}{2} + \beta -\varepsilon};\tag 1.48
$$
$$
\| |p|^{1/2}\delta E_n \tilde{E}_n (H) \| \leq C_\varepsilon
2^{-n+\varepsilon n  -\beta n} |\|V\||_{\frac{3}{2} + \beta -
\varepsilon}, \tag 1.49
$$
with
$$
\| | V \| |_\alpha \equiv \|\langle x \rangle^\alpha V(x)
\|_{L^\infty}.
$$

\endproclaim

%2/9/15

\demo{Proof}
$$
\aligned
&\|\tilde {E}_n \delta E_n\| = \|\delta E_n \tilde{E}_n \|=\\
&= \|\int\hat{E}_n (\lambda) e^{i\lambda|p|} \int^\lambda_0 e^{-is
|p|} V(x) \langle x \rangle^1 \langle x \rangle^{-1}
 |p|^{-1} |p| \tilde{E}_n (H) e^{isH} d s d\lambda \|=\\
&=\| \int \hat{E}(\lambda)\left( \partial_\lambda
e^{i\lambda |p|}\right)
\int^\lambda_0 e^{-is |p|} \frac{1}{|p|} V(x)\langle x \rangle^1
\langle x\rangle^{-1} |p|^{-1} |p| \tilde{E}_n(H) e^{isH} d s d\lambda
\|\\
&\leq \|\int \frac{\partial \hat{E}_n(\lambda)}{\partial \lambda} \int^\lambda_0
e^{-is |p|} \frac{1}{|p|}\frac{1}{\langle x \rangle }
\left( \langle x \rangle^2 V(x)\right)\left(\langle
x\rangle^{-1} |p|^{-1}\right) |p| \tilde{E}_n (H) e^{isH} d s d
\lambda \|\\
&+\|\int\hat{E}_n (\lambda) \frac{1}{|p|}\frac{1}{\langle x \rangle}
\left( \langle x \rangle^2 V(x)\right) \left( \langle x \rangle^{-1}
|p|^{-1}\right) |p|\tilde{E}_n(H) d \lambda \|.\endaligned
\tag 1.50
$$

The second term on the $RHS$ of (1.50) is zero, since
$$
\int^\infty_{-\infty} \hat{E}_n (\lambda) d \lambda = E_n (0) = 0.
$$
The first term on the $RHS$ of (1.50) is bounded by
$$
\int\big| \lambda \frac{\partial \hat{E}_n(\lambda)}{\partial \lambda}
\big| d \lambda \| \frac{1}{|p|} \frac{1}{\langle x \rangle}\|^2
\|\langle x \rangle^2 V(x) \|_{L^\infty} \| |p|\tilde{E}_n(H) \|
$$
$$
\leq C2^{-n} \int\big|\lambda \frac{\partial
\hat{E}_n(\lambda)}{\partial \lambda}  \big| d \lambda \leq C 2^{-n},
$$
since $\lambda \frac{\partial
\hat{E}_n(\lambda)}{\partial \lambda}= \Cal F\left( x E^\prime_n (x)
\right)$ and $x E^\prime_n (x)$ is a $C^\infty_0$ function bounded by
$O(1)$.

This proves (1.47).

The proof of (1.48) follows a similar argument:

For $0\leq  \beta \leq 1:$

$$
\|  |p|^{1/2 + \varepsilon} (\delta E_n) E_n (H) \| =
$$

$$
\|
\int \hat{E}_n (\lambda) \left( \partial^2_\lambda e^{i\lambda |p|}
\right)
$$
$$ \int^\lambda_0 e^{-is |p|} |p|^{-3/2 + \varepsilon} \langle x
\rangle^{-3/2 + \varepsilon} \langle x \rangle^{3/2 - \varepsilon}
V(x) \langle x \rangle^{+\beta} \langle x \rangle^{-\beta}
|p|^{-\beta} |p|^\beta \tilde{E}_n (H) e^{is H} dsd\lambda\|
$$
$$
\leq c \int\big| \lambda \frac{\partial^2 \hat{E}_n
(\lambda)}{\partial \lambda} \big| d \lambda \| V(x) \langle x
\rangle^{3/2 + \beta - \varepsilon} \| 2^{-\beta n} 2^{-n} +
$$
$$
+c\int\big| \frac{\partial \hat{E}_n(\lambda)}{\partial \lambda}
\big| d \lambda 2^{-\beta n} \| \langle x \rangle^{3/2 + \beta -
\varepsilon} V (x) \|_{L^\infty}
$$
$$
+ c \| \int \hat{E}_n (\lambda) |p|^{-3/2 + \varepsilon} \langle x
\rangle^{-3/2 + \varepsilon}\langle x
\rangle^{3/2 - \varepsilon} V(x) \langle x \rangle^\beta\langle x
\rangle^{-\beta} |p|^\beta |p|^{-\beta}  \tilde{E}_n(H) H e^{i\lambda H} d \lambda
\|
$$
$$
\leq C2^{-2n-\beta n} \|V\langle x \rangle^{3/2 +\beta -\varepsilon}
\|_{L^\infty} +(C2^{-n} 2^{-\beta n} +C 2^{-n-\beta n}) \|V(x) \langle
x \rangle^{3/2 + \beta + \varepsilon} \|_{L^\infty}.
$$

\enddemo

%3/2/15 pg18 Feb 27
  Estimate (1.49)
follows by interpolation of (1.47) and (1.48).

$\blacksquare$

Next, we consider the region $\frac {|\lambda|}{Rt} \leq m$.
Now, $\frac {|\lambda|}{Rt} \leq m$ implies  $\frac {|s|}{Rt},
\frac{|u|}{Rt}\leq m$.   Therefore $\quad e^{|s|/Rt}, e^{|u|/Rt} \leq e^m \quad$,
 so, $\quad |p|_s \leq e^m |p| \quad $.
So, we pick up a factor of $2^{-n}$ from $|p|^{1/2}$ factors, and the
integration of $u, s, t $ gives a quantity
$$
    \int_{\frac {|\lambda|}{Rt} \leq m} |\lambda^2 \hat{F}_n (\lambda)
    | d\lambda \leq c 2^{2n}. \tag 1.51
$$
Hence,
$$
    \bigg| \int_{2^n/R} ^T \frac {dt}{R^2t^2}\int \int \tilde{J}(\lambda, s, u) \hat{F}_n(\lambda) d\lambda ds du
    \bigg|
    \leq c \frac {e^m/2}{R} 2^{-n} (2^{2n}  2^{-n/2} 2^{-n (3/2 - \varepsilon)}
$$
%where we gain an extra $2^{-n}$
%from $E_{\bar n} E_n$ and $|p|$
%each.

\smallskip

Here $2^{-n}/R$ comes from the $t$ integration of $(Rt)^2$. A factor of $2^{2n}$ comes form the $\lambda$ integration of
 equation (1.51).
 $e^{m/2}$ is the bound on $e^{u/2Rt}$. $2^{-n/2}$ comes from $|p|^{1/2}E_n(H).$
 $2^{-n(3/2-\varepsilon)}$ comes from $|p|^{1/2}\delta E_n$.
 $\blacksquare$

\medskip

\enddemo
\smallskip
\head {\bf The region $0 \leq t \leq 2^n / R.$} \endhead

\medskip

\proclaim{Proposition B}

 For $\Phi_n$ as in (1.2a), we have:
$$
\int_1^{2^n/R} \|G_n(A/t)E_n\psi(t)\|^2\frac{ dt }{t}\le
 c\|E_n\psi(0)\|^2. \tag 1.52
$$
Here, $G_n$, is a bump function of $A/t$ around $2^{-n}/R.$
\endproclaim
\demo{Proof}

Now we have,
$$
\frac{d}{dt}\left( \psi(t), \Phi_n \psi(t)\right) = \left( \psi(t),
i[H,\Phi_n] \psi(t)\right) + \left(\psi(t), \frac{d\Phi_n}{dt}
\psi(t)\right),
$$
$$
\left(\psi(t),\frac{d\Phi_n(t)}{dt} \psi(t)\right) =
\left(E_n\psi(t), t^{-1}
F_n^{'}(A/t)\left(\frac{-2^n}{R} \frac{A}{t}\right)E_n\psi(t)\right)
$$
$$
\leq -\frac{1}{t}\left(E_n\psi, F_n^{'}(A/t)E_n\psi\right)
\equiv -\frac{1}{t}\left(E_n\psi,\tilde{F}_n^2(A/t)E_n\psi\right),
$$
where,
$$
\align
&|\tilde{F}_n^2(A/t)| \lesssim 1 \quad \text{and} \quad \tilde{F}_n^2 \geq 0 \\
&\tilde{F}_n^2(A/t) \simeq 1 \quad \text{for} \quad \frac{A}{t} \sim R2^{-n} \\
&\tilde{F}_n^2(A/t) = 0 \quad \text{for} \quad \frac{A}{t} \nsim
R2^{-n}.
\endalign
$$
$$
\align
|\left(\psi(t),i[H,\Phi_n]\psi(t) \right) | &=|  \left (\psi(t),i(HE_nF_nE_n - E_nF_nE_nH)\psi(t) \right) | \\
&= | \left(E_n\psi(t),(HE_nF_nE_n - E_nF_nE_nH)E_n\psi(t)\right) | \\
&\leq 2\cdot 2^{-n} ||E_n\psi(t)||^2,
\endalign
$$
 since $ \|HE_n \|\leq 2^{-n}$, and $F_n$ is bounded.
 Hence,
$$
\align
&\int_1^{2^n/R}\biggl\{\Bigl(\psi(t),i[H,\Phi_n]\psi(t)\Bigr) + \Bigl(\psi(t),\frac{d\Phi_n}{dt}\psi(t)\Bigr)\biggr\}dt
 \\
&\leq -\int_1^{2^n/R}\frac{dt}{t}\Bigl\| \tilde{F}_n(A/t)E_n\psi(t)\Bigr\|^2 +
 2\Bigl\|E_n\psi(t)\Bigr\|^2 \int_1^{2^n/R}2^{-n}dt \\
&= - \int_1^{2^n/R} \frac{dt}{t} \Bigl\| \tilde{F}_n (A/t) E_n \psi(t) \Bigr\|^2 +
 \frac{2}{R} \Bigl\|E_n\psi(t)\Bigr\|^2.
\endalign
$$

Therefore,
$$
\align
\int_1 ^{2^n/R} \frac{dt}{t} &\Bigl\| \tilde{F}_n (A/t) E_n \psi
(t)\Bigr\|^2 + \Bigl( \psi (2^n/R), E_n F(A\geq 1) E_n \psi(2^n/R) \Bigr)\\
&\leq \Bigl( E_n \psi(1), F_n(A,t=1) E_n \psi(1) \Bigr) +
\frac{2}{R} \Bigl\| E_n \psi(t) \Bigr\|^2.
\endalign
$$

%$\blacksquare$
since $F_n(A, t = 2^n/R) = F(\frac{A}{2^{-n}(2^n/R)}\geq R) = F(A R\geq
R) = (A\geq 1)$.

\enddemo
\smallskip
 \centerline {\bf Improved Decay}
\medskip
\proclaim {Proposition C}

Under the assumptions as the above Propositions (A,B), we have the
following propagation estimate,
$$
\align
 \int_{2^n/R}^T \| |\frac{A}{t}|^{1/2}   F_n\left(\frac{A}{Rt2^{-n}}>1\right
)E_n\psi(t)\|^2& \frac{dt}{t} \\ +
<\psi(T),E_n\frac{A}{T}F_n\left(\frac{A}{RT2^{-n}}>1\right
)E_n\psi(T)> & \le c(R)\|E_n |A|^{1/2}F(A\geq R2^{-n})   E_n\psi(0)\|^2 \tag 1.53
\endalign
$$
for all $ R \geq 1 + \varepsilon $.
\endproclaim

\demo{Proof}
 We can use $E_n \frac{A}{t} F_n (A/t) E_n = \Phi_n $ .
The estimate of $R_2(A/t)$ is done as before, so for $t\geq 2^n$
, it is done as before.The main change is that now $\hat F(\lambda)$
is replaced by $\partial_{\lambda} \hat F(\lambda)$. The first term in the commutator expansion of
$i[H,\Phi_n],$ has an extra, positive term, which is however
integrable over time by the previous propositions (A,B), since it is
supported (in phase space) on the support of $F_n^{\prime}.$
%$\blacksquare$
\enddemo

$$
\aligned
& i[H, E_n \frac{A}{t} F_n (A/t) E_n] + E_n\partial_t \frac{A}{t} F_n
(A/t)E_n = \\
& = E_n( -\frac{A}{t^2} F_n - \frac{A^2}{t^3} F^\prime_n (A/t))E_n +
E_n i[V(x), \frac{A}{t} F_n (A/t)] E_n\\
& + E_n i[|p|, \frac{A}{t}]F_n(A/t) E_n + E_n \frac{A}{t} i[|p|,
F_n(A/t)]E_n\endaligned\tag 1.54
$$
$$
\leq - \frac{1}{t} E_n \left\{ 2^{-n} R - |p| + x\cdot \nabla
V\right\} F_n(A/t) E_n \tag 1.55
$$
$$
+\frac{1}{t} E_n \left\{\frac{A}{t} F'_n (A/t) [|p| - x\cdot \nabla
V] + \frac{A}{t} R_2 (A/t) \right\} E_n \tag 1.56
$$
$$
- E_n \frac{1}{t} 2^{-n} R \left(\frac{A}{t} F^\prime_n\right) E_n.
\tag 1.57
$$

Now,
$$
E_n(-|p| + x\cdot \nabla V) = E_n \tilde{E}_n (H) (-|p| + x\cdot
\nabla V) = E_n (2^{-n} + 0(1) 2^{-n})\tag 1.58
$$
$$
= - \frac{1}{t} E_n \left\{2^{-n} (R-1-0(1)\right\} F_n
(A/t) E_n.\tag 1.59
$$
$$
\leq E_n(2^{-n}R) \frac{1}{t} \tilde{F}_n E_n = 2^{-n} R O
(L^1(dt))\tag 1.60
$$

by our previous propagation estimates.

Since $\frac{A}{t} F^\prime_n (A/t) = \tilde{F}_n (A/t)$, and $(|p|-
x\cdot \nabla V) E_n = O(2^{-n})$ it follows that
$$
 O(2^{-n})E_n \frac{1}{t} \tilde F_n (A/t)E_n + E_n \frac{A}{t} R_2 (A/t)
E_n = O(2^{-n}) O ( L^1(dt) + E_n \frac{A}{t} R_2 (A/t) E_n.
\tag 1.61
$$
The term $\frac{A}{t} R_2 (A/t)$ is the same as the $R_2(A/t)$ we
treated before, except, that through integration by parts,
$\frac{A}{t}$ factor is absorbed into the $R_2(A/t)$ term, by
replacing $\hat{F}_n(\lambda)$ by
$\partial_\lambda\hat{F}_n(\lambda)$.

Since $\partial_\lambda\hat{F}_n(\lambda) = \Cal{F}(xF_n(x))$, and
$xF_n(x) = O(2^{-n}R)$ we conclude that this last term is also of
order
$$
2^{-n} R O(L^1(dt)).
$$

collecting all the estimates on (1.54), we arrive at (1.53), which proves Proposition C.

$\square$

\remark{Remark}

Since
$$|(E_n \psi(1) , A E_n\psi(1) | \leq C\| |p| E_n H H^{-1}
\psi(1) \|\left[ \|[x, E_n]\psi (1) \| + \| E_n x\psi(1) \|\right]
$$
$$
 \leq C
2^{-2n} \| H^{-1}P_c(H) \langle x \rangle^{-1} \| \| \langle x \rangle
\psi(1) \left[ O(1) \| \psi\| + \| x \psi(1) \|\right] = O(2^{-2n})\|<x>\psi(1)\|^2_2
$$
then, by factoring out a factor of $2^{-n} R$, we get a bound on
$$
\int^T_1 \| F_n (A. t) E_n \psi(t) \|^2 \frac{dt}{t} \leq C\left(
\|E_n\psi\|^2 + 2^{-2n} \| | \psi |  \|^2_1\right)
$$
$$
 \| |\psi| \|_1 = \|
\langle x \rangle \psi (1) \|_{L^2}.
$$

\endremark

%/27/15

For $0\leq t\leq 2^n$, since $\frac{A}{t}F_n$ is not bounded, the
proof is different.

\proclaim{Proposition D}
$$
\align  \int_1^{2^n/R} \| \sqrt(A/t)   F_n\left(\frac{A}{Rt2^{-n}}>1\right
)E_n\psi(t)\|^2& \frac{dt}{t} \\+
<\psi(2^n/R),E_n\frac{A}{2^n/R}F_n\left(A>1\right )E_n\psi(t)> & \le
c(R)\|E_n <A>^{1/2}E_n\psi(1)\|^2, \tag 1.62
\endalign
$$
for all $ R> 1 + \varepsilon $.
\endproclaim

\demo{Proof}
$\partial_t \left( \frac{A}{t} F_n\right) = - \frac{A}{t^2} F_n -
\frac{A}{Rt^2 2^{-n}} F^\prime_n \leq 0$.  This gives the leading
term, in the Heisenberg derivative of the Propagation observable
$\frac{A}{t} F_n (A/t)$.

%2/27/15

 We write,
$$
\frac{A}{t}F_n = F_{n,1} + F_{n,2},\tag 1.63
$$
where,
$$
F_{n,1} =i \int e^{-\lambda ^2} (\partial_\lambda \hat{F}_n(\lambda))
e^{-i(A/t)\lambda} d\lambda ,\tag 1.64
$$
and
$$  \quad F_{n,2} \equiv (A/t)F_n - F_{n,1}.\tag 1.65
 $$
Then, $F_{n,2}$ is bounded, and the previous proof applies, while
$F_{n,1} \equiv \frac{A}{t}G_n$, with $ G_n$ smooth, approaching a
constant at infinity. Then,
$$
\align
i[H,F_{n,1}] &= \frac{1}{t}i[H,A]G_n + \frac{A}{t}i[H,G_n] \\
&= \frac{1}{t}\bigl[|p| - x \cdot  \nabla V \bigr]G_n +\frac{A}{t}\int \hat{G}_n(\lambda) e^{-i\lambda A/t}
\int_0^\lambda e^{isA/t} [H,A/t] e^{-isA/t} ds d\lambda. \tag 1.66
\endalign
$$

The first term is bounded by $  2^{-n}\|G_n\|/t  \quad $ on support
of $E_n$, and therefore its integral over$  1 \leq t \leq2^n$
 is bounded by $O(1)$.
 The second term is,
$$
\align
\frac{1}{t}\int\partial_\lambda \Bigl(\lambda \hat{G}_n(\lambda)\Bigr) e^{-i\lambda A/t} &\frac{1}{\lambda}
 \int_0^\lambda e^{-s/t} \Bigl[ |p|+\tilde {V}(x,s)\Bigr]ds d\lambda \\
& - \frac{1}{t}\int \hat{G}_n(\lambda) e^{-i\lambda A/t}
\frac{1}{\lambda} \int_0^\lambda \Bigl[e^{-s/t} |p|+ {V}(x
e^{s/t})\Bigr]ds d\lambda.\tag 1.67
\endalign
$$
$$ \frac{1}{\lambda}\int_0^\lambda e^{-s/t} ds = -\frac{t}{\lambda}
e^{-s/t} |_0^\lambda = \frac{t}{\lambda} \bigl( e^{-\lambda /t} -1
\bigr)= -1 + \frac{1}{2} \left( \frac{\lambda}{t}\right) + O
\left(\left(\frac{\lambda}{t}\right)^2\right)
$$
for  $\frac{\lambda}{t} \ll 1$.

So the first term of (1.66)
contributes $| \int \hat{G}_n (\lambda) e^{i\lambda A/t} d\lambda|
\frac{1}{t} 2^{-n} = O(\frac{1}{t}) 2^{-n},$
and the second term is bounded by
$$\int|\partial_\lambda ( \lambda
\hat{G}_n(\lambda))| d\lambda /t +t^{-1}  \int _{|\lambda |\geq 1}
|\hat{G}_n (\lambda)| e^{-\lambda/t} / d \lambda $$.

%$\blacksquare$
%\enddemo

%3/6/15 red page 35

We use
$$
\int | \hat{G}_n (\lambda) |d\lambda \leq \int | \hat{F}_n
(\lambda) e^{-\lambda^2} | +Erf (\lambda|\partial_\lambda
\hat{F}_n (\lambda)|) d\lambda = O (2^{-n});
$$
$Erf$ stands for the Error function.
Furthermore, there is a factor of $2^{-n}$,coming from $|p|+V$:
Here $ \tilde{V} (x, s)=e^{isA/t}\tilde V(x)e^{-isA/t} $.
$
\tilde{V} (x, s) = \tilde{V} (e^{-s/t} x)$, and for $|\frac{s}{t}|\leq
1$, we use $|p| + \tilde{V} (x, s) \leq C |p|$.

\enddemo
$\blacksquare$

%red page 36 3/6

 \head {Section 2. Auxiliary Identities and Inequalities}
\endhead
\medskip
\proclaim {Lemma 2.1}
Assume $H=|p|+V$ and $|V| <
\frac{1}{2r}$. Then,
$$
\Bigl\||p|E_{I_n}(H)f\Bigr\|+\Bigl\|V E_{I_n}(H)f\Bigr\| \leq c2^{-n}
\Bigl\|E_{I_n} (H) f\Bigr\| .\tag 2.1
$$
\endproclaim
\demo{Proof}
Let $\|f\|=1$.
$$
\align
\Bigl(f,E_{I_n}(H)\bigl(|p|^2+V^2\bigr)E_{I_n}(H)f\Bigr) &= \Bigl(f,E_{I_n}(H)\{H^2-V|p|-|p|V\}E_{I_n}(H)f\Bigr) \\
& \leq
2^{-2n}\Bigl\|E_{I_n}(H)f\Bigr\|^2+2\Bigl\|VE_{I_n}(H)f\Bigr\|\Bigl\||p|E_{I_n}(H)f\Bigr\|.
\endalign
$$
Let $a=\Bigl\||p|E_{I_n}(H)f\Bigr\| \quad b=\Bigl\|VE_{I_n}(H)f\Bigr\|$.
 Then, $\quad b<(1-\delta)\Bigl\||p|E_{I_n}(H)f\Bigr\|$ by
the uncertainty inequality(in 3-dimensions or higher), and so, \quad
$$
\align
\Bigl(f,E_{I_n}(H)\bigl(|p|^2+V^2\bigr)E_{I_n}(H)f\Bigr) &= (a^2+b^2-2ab) + 2ab\\
&\geq\delta ^2\Bigl\||p|E_{I_n}(H)f\Bigr\|^2 + 2ab.
\endalign
$$
It follows  that, $\quad \delta ^2 \Bigl\||p|E_{I_n}(H)f\Bigr\|^2 \leq
2^{-2n} \Bigl\|E_{I_n}(H)f\Bigr\|^2$ .$\blacksquare$
\newline
\enddemo
\proclaim{Proposition 2.2} If $|p| \lesssim H$,  then (2.1)
holds.
\endproclaim
\demo{Proof}
 When $|p| \leq mH$, we have that, by
the spectral theorem,
$$
\frac{1}{H}\leq\frac{m}{|p|} \Longrightarrow \Bigl\|\frac{1}{H^{1/2}}f
\Bigr\|_{L^2} \leq m^{1/2}\Bigl\|\frac{1}{|p|^{1/2}}f\Bigr\| .\tag 2.2
$$
Hence,
$$
V(x)\frac{1}{H} = V(x)\left(\frac{1}{|H|} -
\frac{1}{|p|}\right) + V(x)\frac{1}{|p|}.
$$
$$
\Bigl\|V(x)\frac{1}{|p|}\Bigr\| = \Bigl\|V(x)r\frac{1}{r}\frac{1}{|p|}\Bigr\|
\leq \Bigl\|V(x)\frac{1}{r}\Bigr\|_\infty
2\Bigl\||p|\frac{1}{|p|}\Bigr\| <\infty,
$$
$$
\align
V(x)\left(\frac{1}{H}-\frac{1}{|p|}\right) &=
-V(x)\frac{1}{H}V(x)\frac{1}{|p|} \\
&= -V(x)\frac{1}{H}\frac{1}{\sqrt{r}}   \sqrt{r} V(x)r\frac{1}{r}\frac{1}{|p|} \\
&= -V(x)\sqrt{r}\frac{1}{\sqrt{r}}\frac{1}{H}\frac{1}{\sqrt{r}}r^{3/2}
V(x)\frac{1}{r}\frac{1}{|p|}.
\endalign
$$
Since,
$$
\frac{1}{\sqrt{r}}\frac{1}{H}\frac{1}{\sqrt{r}}=\frac{1}{\sqrt{r}}\frac{1}{H^{1/2}}\frac{1}{H^{1/2}}\frac{1}{\sqrt{r}}
=\left(\frac{1}{H^{1/2}}\frac{1}{\sqrt{r}}\right)^*\left(\frac{1}{H^{1/2}}\frac{1}{\sqrt{r}}\right),
$$
we get that,
$$
\align
\Bigl\|V(x)\left(\frac{1}{H}-\frac{1}{|p|}\right)\Bigr\|
&\leq \Bigl\|\frac{1}{H^{1/2}}\frac{1}{\sqrt{r}}\sqrt{r}V(x)\Bigr\|\Bigl\|\frac{1}{H^{1/2}}
\frac{1}{\sqrt{r}}r^{3/2}V(x)\frac{1}{r}\frac{1}{|p|}\Bigr\| \\
&\leq
m\Bigl\|\frac{1}{|p|^{1/2}}\frac{1}{\sqrt{r}}\bigl(\sqrt{r}V(x)\bigr)\Bigr\|\Bigl\|\frac{1}{|p|^{1/2}}\frac{1}{\sqrt{r}}(r^{3/2}
V)
\Bigr\|\Bigl\|\frac{1}{r}\frac{1}{|p|}\Bigr\| \\
&< mC_V,
\endalign
$$
where $C_V$ is a constant, depending on the $L^\infty$ norm of $V,
\sqrt{r}V,r^{3/2} V$.\newline
We therefore conclude that,
$$
|p|\frac{1}{H} = \bigl(|p|+V-V\bigr)\frac{1}{H} = H\frac{1}{H}-V\frac{1}{H}=1-V\frac{1}{H}
$$
is also bounded.
Finally, we have that,
$$
|p|E_{I_n}(H)=|p|\frac{1}{H}HE_{I_n}(H)=O(1)2^{-n}E_{I_n}(H).\quad
\quad \blacksquare
$$
\enddemo
Now we prove some useful identities.
 \proclaim {Lemma 2.3 } Let the dimension be 3, and denote as before by $A$ the dilation generator, and by $L^2$ the Laplacian on the unit sphere.
 Then the following identities and an inequality hold:
\medskip
\noindent (i)\quad $r^2(-\Delta)=A^2+L^2-2iA-3/4, L^2 $ stands for the
 Laplacian on the unit sphere.
\newline
(ii) \quad $e^{i\lambda A}|p|^\alpha e^{-i\lambda A}=e^{-\alpha
\lambda}|p|^\alpha$ ; \quad $A\equiv
\frac{1}{i}(r\partial_r+\frac{3}{2}) = (x\cdot p+p \cdot x)/ 2$
\newline (iii)  \quad
$e^{i\lambda A} r^\alpha e^{-i\lambda A}=e^{\alpha \lambda} r^\alpha
. $ \newline (iv)    \quad
$\frac{1}{|p|}A\frac{1}{r}=\frac{1}{|p|}\Bigl(i|p|\frac{\partial}{\partial
|p|}-\frac{3i}{2}\Bigr)\frac{1}{r}=\frac{1}{|p|}\Bigl(ip\cdot
\frac{\partial}{\partial p} -\frac{3i}{2}\Big)\frac{1}{r}=O(1).$
\newline (v) \quad $\Bigl\|\frac{1}{|p|}\frac{1}{r}\psi\Bigr\| \leq
2\|\psi \|. $\endproclaim

  \demo{Proof}
  \newline (i) \quad
$r^2(-\Delta)=r^2\Bigl(-\partial_r^2-\frac{2}{r}\partial_r
+\frac{L^2}{r^2}\Bigr)=-r^2\partial_r^2-2r\partial_r+L^2$
\newline
$=-2r\partial_r+2r\partial_r-r\partial_r^2r+L^2=-r\partial_r^2r+L^2,$
\newline
$A^2=-\frac{1}{4}\Bigl(+2r\partial_r + 3\Bigr)^2=(-4r\partial_rr\partial_r-9-12r\partial_r)\frac{1}{4}$
\newline
$=r\partial_r-r\partial_r^2r-\frac{9}{4}-3r\partial_r=-r\partial_r^2r-2r\partial_r-\frac{9}{4}$
;\newline so,\quad
$r^2(-\Delta)=-r\partial_r^2r+L^2=A^2+\frac{9}{4}+L^2 +2 iA-3/4
=A^2+L^2- 2 iA-3.$
\newline
(ii) \quad $\partial_\lambda\Bigl(e^{i\lambda A}|p|^\alpha
e^{-i\lambda A}\Bigr)=e^{i\lambda A}
i\Bigl[A,|p|^\alpha\Bigr]e^{-i\lambda A}=-\alpha e^{i\lambda A}
|p|^\alpha e^{-i\lambda A}$\newline $ \Rightarrow  e^{i\lambda A}
|p|^\alpha e^{-i\lambda A}=e^{-\lambda\alpha}|p|^\alpha .$ \newline
(iii)\quad As in(ii), but now, $i[A,r^\alpha]=\alpha r^\alpha .$
\newline (iv)\quad Follows from (v). \newline (v)\quad
$\Bigl\|\frac{1}{|p|}\frac{1}{r}\psi\Bigr\|_{L^2}
\leq2\Bigl\|r\frac{1}{r}\psi\Bigr\|_{L^2} = 2\|\psi\|$
\newline
by the
uncertainty inequality in 3 dimensions.$\blacksquare$
\enddemo
\proclaim {Proposition 2.4}
Let $n\geq 1$, the hamiltonian $H$ and the momentum operator $p$ are as before.
Then, for all $n$ the following estimates hold:

a) $\quad E_n^3(H) = E_n(H) E_n(|p|) E_n(H) + E_n(H) \delta^{-1}
O(2^{-n}) E_n(H).$

b) $\quad E_{\bar{n}}(|p|) E_n(H) = E_{\bar{n}}(|p|) O(2^{-n})
E_n(H), \quad \quad \bar{n} \neq n .$ \newline  For $\quad \eta > \ln
2,$

c) $\quad E_n (|p|) e^{i\eta A} E_n(H) = E_n(|p|) O(2^{-n}) E_n(H).$

d) $\quad \bar{E}_n (|p|) E_n(H) = \bar{E}_n(|p|) O(2^{-n}n) E_n(H).$
\endproclaim

\demo{Proof}

Part(a).  Let $E_n (H) \equiv E_n$.

$E^3_n(H) = -E_n(E_n(|p|)-E_n)E_n + E_n(H)E_n(|p|) E_n(H),$

$E_n(|p|) - E_n(H) = \int \hat{E}_n(\lambda)\left(e^{i\lambda |p|} - e^{i\lambda H}\right) d\lambda$

$\quad \quad \quad \quad \quad \quad \quad \quad= \int \hat{E}_n(\lambda) e^{i\lambda H}
 \int_0^\lambda e^{-is H} (-i) V e^{is |p|} ds d\lambda .$

Therefore, using
$$
    |V| \leq c<x>^{-2},
$$
we get:
$ \left|\left( \phi, E_n\left[ E_n(|p|) - E_n(H)\right]E_n \psi \right) \right|$

$ \leq c \int \left\| <x>^{-1} e^{-is H} e^{i\lambda H} E_n \phi\right\| |\hat{E}_n(\lambda)\lambda|
 \frac{1}{\lambda} \int_0 ^\lambda \left\|<x>^{-1} e^{-is |p|} E_n \psi \right\| ds d\lambda$

$ \leq c\delta^{-1} \left\| HE_n\phi\right\| \int|\lambda \hat{E}_n(\lambda)|d\lambda \left\| |p|E_n \psi\right\|$

$ \leq c\delta^{-2} 2^{-n}2^n 2^{-n}\|\phi\| \|\psi\| =
c\delta^{-2}2^{-n}\|\phi\| \|\psi\|,$

where we used that,
$$\quad \quad <x>^{-1} \leq c|p| \leq c\delta^{-1}H \quad $$
in dimension three or higher, and
\newline
proposition(2.2).

Notice that,
\quad $\| |p| E_n\psi\| \leq c\delta^{-1}\|HE_n\psi\| \leq
c\delta^{-1} 2^{-n}\|E_n\psi \|$, \quad by Lemma (2.3).

Part(b).

$\left|\left(\phi,E_{\bar{n}} (|p|)E_n(H)\psi\right)\right| =
|\left(\phi,E_{\bar{n}}(|p|)\left[\tilde{E}_{\bar{n}}(|p|)-\tilde{E}_{\bar{n}}(H)\right]E_n(H)\psi\right)|$

$\leq  c\delta^{-1} \| HE_n\psi\| \||p|E_{\bar{n}}(|p|)\phi\|
\int|\hat{E}_{\bar{n}}\lambda|d\lambda ,$

by the proof of Part(a).
The last expression is therefore bounded by,
$$
    c \delta^{-1}2^{-n}\|\psi\|2^{-{\bar{n}}}\|\phi\|2^{\bar{n}} =
    c\delta^{-1}2^{-n}\|\phi\| \|\psi\|= O(2^{-n}).
$$

Part(c) follows from Part(b), since, for $|\lambda| > \ln 2$:

$$
\align
E_n(|p|)e^{i\lambda A}E_n &= E_n(|p|)e^{i\lambda A} \tilde{E}_n(p) E_n \\
                       &+ E_n (p) e^{i\lambda A} \bar{E}_n(p) E_n\\
                    &= 0 + O( \bar{E} (p) E_n) =O (n 2^{-n})\text{ from (d)}.
\endalign
$$

$$
\align
\|E_n(|p|)e^{i\lambda A}\| &=\| E_n(|p|)e^{i\lambda A} E_n(e^{-\lambda} |p|)\| \\
                        &\lesssim \| E_n(|p|) e^{i\lambda A} \sum_{|\bar n |\leq M} E_{\bar{n}} (|p|)\|\\
                        & \text {with} \quad \bar n \neq n, M<
                        \infty .
\endalign
$$

Part(d) follows from part(b), since the domain of $|p|$, in the
support of $\bar{E}_n(|p|)$, is covered by $n$ dyadic intervals from
$[2^{-n},1]$:
$$
\bar{E}_n(p) = 1-E_n(p) = \sum_{\bar{n} \neq n} E_{\bar{n}}(p) = E (|p| >1
) + \sum_{j=1}^{M\leq n} E_{n_j}.
$$
$\blacksquare$

%3/6/15
\enddemo

\proclaim{Proposition 2.5}
Assume that \quad $H\equiv|p|+V(x)$ \quad has no bound states or zero energy resonances,
and that $V(x)$ vanishes faster than $r^{-2}$ at infinity, and is sufficiently regular.
The dimension is 3. Then, for some $0<m<\infty$,
$$
H\geq m|p|. \tag 2.3
$$
\endproclaim
\demo{Proof}  Since $H$ has no bound states, and $V(x)
\rightarrow 0$ at infinity, $H\geq0$. If we now make a small
perturbation, $H\rightarrow H_\epsilon = H + \epsilon V,$  then,
since $H$ has no zero energy resonances, $H_\epsilon$ has no bound
states, for $\epsilon$ sufficiently small. Hence $H_\epsilon \geq
0$. \newline But then,
$$
    H=\frac{1}{1+\epsilon}\Bigl(|p|+(1+\epsilon) V \Bigr)+\Bigl(1-\frac{1}{1+\epsilon}\Bigr)
    |p|=\frac{1}{1+\epsilon}H_\epsilon +m|p|\geq m|p|.\blacksquare
$$
\enddemo

\proclaim{Lemma 2.6}
 For \quad $H=|p|+V$, we have that \quad
$|p|\geq \delta H$, for dimension$=3$, provided, \newline
$|V|\leq\bar{c}/r.$   Moreover : $E_{I_n} (H) |p| E_{I_n} (H) \geq
\delta E_{I_n} (H) H E_{I_n} (H) \geq \delta (\text{inf} I_n)\cdot
E_{I_n} (H)$.

\endproclaim
\demo{Proof}
 For $ \delta > 0$ sufficiently small
$$
\delta H=\delta|p|+\delta V \leq \delta|p| +\delta c |p| =
\delta(1+c)|p| \leq |p| , \delta (1+c) < 1, \tag 2.4
$$
where we used that for $|V|\leq\bar{c}|x|^{-1}$, we have $|V|\leq
c|p|$ in $3$ dimensions.\newline So, in particular, we have,
$$
    E_{I_n}(H)|p|E_{I_n}(H) \geq E_{I_n}(H)\delta HE_{I_n}(H)\geq \delta \inf I_n E_{I_n}(H). \tag 2.5
$$
$\blacksquare$
\enddemo
\proclaim{Proposition 2.7}
Suppose, as before, that $H\equiv |p|+V(x), $ the dimension is 3,
and that $V(x)$ is sufficiently regular, and vanishes faster than $r^{-2}$ at infinity.
 Suppose, moreover, that $H$ has no zero energy resonances, and no zero energy bound states. \newline
Then,
$$
    P_c(H)HP_c(H) \geq P_c(H)\delta |p| P_c(H),\quad \text{for some}\quad \delta>0 .\tag 2.6
$$
\endproclaim
\demo{Proof}
We have that,
$$
    P_c(H)\Bigl(|p|+V\Bigr)P_c(H)\geq 0.
$$
Add a small perturbation $\epsilon V$ to $H$: \quad $H_\epsilon
\equiv H+\epsilon V=|p|+(1+\epsilon)V .$ \newline Then, for $\epsilon$
sufficiently small, no new bound states are created. \newline Hence,
$$
    f(H_\epsilon\geq-\epsilon_0) \bigl(|p|+V+\epsilon V\bigr) f(H_\epsilon\geq -\epsilon_0)\geq 0 ,\tag 2.7
$$
for some $\epsilon_0 >0$, and $f$ is a smooth characteristic function of the interval $[-\epsilon_0 , \infty]$. \newline
Then,
$$
    P_c(H) H P_c(H) =
    P_c(H)\Bigl(\frac{1}{1+\epsilon}H_\epsilon+\epsilon_1
    |p|\Bigr)P_c(H), \varepsilon_1 \equiv 1 - \frac{1}{1+\varepsilon}
    = \frac{\varepsilon}{1+\varepsilon}
$$
would imply that:
$$
    P_c(H)HP_c(H) \geq \epsilon_1 P_c(H)|p|P_c(H),
$$
if we can prove that for some $\epsilon_2 < \epsilon_1$,
$$
    P_c(H)H_\epsilon P_c(H)\geq P_c(-\epsilon_2|p|)P_c .\tag 2.8
$$
So we want to use (2.7) to prove (2.8).
$$
\align
P_c (H)H_\epsilon P_c(H)&=P_c(H)f(H_\epsilon\leq-\epsilon_0)H_\epsilon P_c(H)
+P_c(H)f(H_\epsilon\geq -\epsilon_0)H_\epsilon P_c(H) \\
&\geq P_c(H)f(H_\epsilon \leq -\epsilon_0)H_\epsilon P_c(H), \tag 2.9
\endalign
$$
by (2.7).
Now,
$$
\align
P_c(H)f(H_\epsilon \leq-\epsilon_0)&=P_c(H)\Bigl[f(H_\epsilon \leq -\epsilon_0) - f(H\leq -\epsilon_0)\Bigr] \\
&= P_c(H)\int\hat{f}(\lambda)e^{+i\lambda H}i\int_0^\lambda e^{-isH}
\epsilon Ve^{isH_\epsilon}ds d\lambda \\
&=P_c(H)O(\epsilon V)\int|\lambda  \hat{f}(\lambda)| d\lambda \\
&=P_c(H) O(\epsilon V) O(\epsilon_0^{-1}),\tag 2.10
\endalign
$$
so,
$$
\align
&P_c(H)f(H_\epsilon\leq -\epsilon_0)H_\epsilon P_c(H) \\
& \quad = P_c(H)O(\epsilon V/\epsilon_0)\tilde{f}(H_\epsilon \leq-\epsilon_0)H_\epsilon
 \Bigl(\tilde{g}(H_\epsilon\leq-\epsilon_0)+\bar{g}(H_\epsilon \leq -\epsilon_0)\Bigr) P_c(H) \\
& \quad = P_c(H)O(\epsilon V/\epsilon_0)\tilde{f}(H_\epsilon)H_\epsilon \tilde g(H_\epsilon \leq -\epsilon_0)P_c(H) \\
& \quad = P_c(H)O(\epsilon
V/\epsilon_0)\tilde{f}(H_\epsilon)H_\epsilon O(\epsilon
V/\epsilon_0)P_c (H) ,\tag 2.11
\endalign
$$
since \quad $\tilde{f}(H_\epsilon\leq
-\epsilon_0)\bar{g}(H_\epsilon\leq -\epsilon_0) = 0.$
$(\tilde{g}+\bar{g}\equiv 1, \tilde{g} \tilde{f}=\tilde{f})$.
 The last term can be bounded by:

$$
   c P_c(H)\epsilon^2(\epsilon_0^{-2} H_{\epsilon})P_c(H)
     \leq c \epsilon_0^{-2} P_c \epsilon^2|p|P_c(H) (H),\tag 2.12
$$

$$
1/\epsilon_0
\text{coming from :}
\int |\lambda \hat{f}(\lambda)|d\lambda = O(1/\epsilon_0).
$$
$\epsilon_0$ is basically the distance of zero to the largest
(negative) e.v. of $H_\epsilon$ , and $\epsilon$ is arbitrarily
small, so \quad $\epsilon^2 / \epsilon_0^2 \leq O(\epsilon^2)$.
Hence, (2.9) - (2.12) imply (2.8). \quad $\blacksquare$
\enddemo

\head  { Section 3. Maximal velocity bound} \endhead
 We begin with
estimating, for $a>1$,
$$
    E_n(H) F_a \left( \frac{|x|}{t} > a \right) E_n(H) \equiv E_n F_a E_n .\tag 3.1
$$
$$
    E_n F_a E_n = E_n F_a (F_n(A/t) + \bar {F}_n(A/t)) E_n, \tag 3.2
$$
where,
$$
    F_n(A/t) \equiv F_n(\frac{A}{t} > R2^{-n}) = 1 - \bar{F}_n ,\tag 3.3
$$
$$
    1 < R < a, \tag 3.4
$$
and such that
$$
 \bar{F}_n(b\leq R2^{-n})\equiv  \bar{F}_n(b)\quad \text {satisfies}\quad\bar
 F_n(2^{-n}b)F_a(b) = 0. \tag 3.5
$$

Since, by the propagation estimates,
$$
    \|F_n (A/t) E_n (H) \psi(t) \| \leq o(1) \|E_n\psi_0\|,
$$
with $o(1) \rightarrow 0$  as $t \rightarrow \infty$, we need to
control $E_n F_a \bar {F}_n E_n$ by a decaying function of $t$, in
order to prove the maximal velocity bound, on the energy shell
$I_n.$

\proclaim{Proposition 3.1}
$$
\|E_n (H)F_a( \frac{r}{t})\bar{F}_n(A/t)\|=
$$
$$
\|E_n (H) |p|^{-1} |p| F_a( \frac{r}{t})\bar{F}_n(A/t)\|\leq
c2^n/t.
$$
\endproclaim

%3/12/15 typing

\demo{Proof}

$$
\align
E_n F_a \bar {F}_n
&= E_n (|p|^{-1} |p|^{-1}\Delta) r^2 F_a r^{-2} \bar {F}_n \\
&= E_n |p|^{-1} |p|^{-1} (A^2 + cA + c_1) (r^{-2} F_a) \bar {F}_n  \\
&= c_2 E_n |p|^{-1} (|p|^{-1} A r^{-1})(A+c) r^{-1} F_a \bar {F}_n  \\
& \quad + c_3 E_n |p|^{-1} (|p|^{-1} r^{-1}) r^{-1} F_a \bar {F}_n \equiv B_1 + B_2. \tag 3.6
\endalign
$$
where in the third equality we used that $[A, r^{-1}] = -ir^{-1}.$

$$
B_2 = E_n O(2^n)O_0(1)c_3
\frac{1}{at}\left[\left(\frac{ta}{r}\right)F_a\right]\bar{F}_n,
\tag 3.7
$$
where we used that,
$$
E_n |p|^{-1} = O (2^{+n}) \quad \text{and} \quad |p|^{-1}r^{-1} =
O_0(1).
$$
Since \quad $\frac{ta}{r} F_a = O(1) , \quad B_2 = O(2^n/at),$
$$
\align
B_1 &= E_n O(2^n) O_1(1) c \frac{1}{at} \left( \frac{ta}{r} F_a\right) \bar {F}_n  \\
&\quad + E_n O(2^n) O_1(1) \frac{1}{at} \left(\frac{ta}{r}F_a \right) Rt (Rt)^{-1} A \bar {F}_n \\
&\quad + E_n O(2^n) O_1(1) \frac{1}{at} \left[A, \left(\frac {at}{r}
F_a \right) \right] \bar {F}_n,
\endalign
$$
where \quad
$$
O_1(1) \equiv |p|^{-1}( A-3/2\imath)
r^{-1}=|p|^{-1}p\cdot xr^{-1}=\sum_{k=1}^3
\frac{p_k}{|p|}\frac{r_k}{r}.
$$
Therefore,
$$
\align
B_1 &= E_n O(2^n / at) \bar{F}_n E_n + E_nO(2^n)O_1(1)\left(\frac{ta}{r}F_a\right) \frac{R}{a}O(2^{-n})\bar{F}_n\\
&= O(2^n/at) + O\left(\frac{R}{a}\right).\text{ We used } \frac{ta}{r}
F_a = O(1), \frac{A}{Rt} \bar{F}_n \lesssim O(2^{-n}). \tag 3.8
\endalign
$$
The key to this computation is the repeated use, as we do below, of the following:
$$
\align
E_n &= E_n|p|^{-1}|p|^{-1}(A^2+cA+c_1) r^{-2} \\
&= c_2 E_n|p|^{-1}(|p|^{-1}Ar^{-1})(A+c^{'})r^{-1} + c_3
E_n|p|^{-1}(|p|^{-1}r^{-1})r^{-1}\\
&=E_n|p|^{-1}O_1(1)(A+c^{\prime})r^{-1}+c_3
E_n|p|^{-1}(|p|^{-1}r^{-1})r^{-1}\\
&+c_4E_n|p|^{-1}(|p|^{-1}r^{-1}A)r^{-1}. \tag 3.9
\endalign
$$
We apply it again, to the \quad $ O\left(\frac{R}{a}\right)$ \quad
term:$ (F^{[k]}(\lambda)\equiv \lambda^k F(\lambda ));$
$$
\align
O(R/a) \sim &\frac{R}{a} E_n O_1(1)F_a^{[-1]}\bar{F}_n^{[1]} \\
&= \frac{R}{a}E_n|p|^{-1}O_1(1)(A+c^{'})r^{-1}O_1(1)F_a^{[-1]}\bar{F}_n^{[1]} \\
& \quad + \frac{R}{a}E_n|p|^{-1}O_0(1)r^{-1}O_1(1)F_a^{[-1]}
\bar{F}_n^{[1]}. \tag 3.10
\endalign
$$
Now, the important observation is that,
$$
 r^{-1}|p|^{-1}Ar^{-1} = (r^{-1}|p|^{-1}A)r^{-1}
$$
and
$$
r^{-1}O_1(1)
=O_1(1)r^{-1}-r^{-1}O(|p|^{-1})r^{-1}=O_1(1)r^{-1}+O(1)r^{-1}
$$
$$
\quad
[A,O_1(1)] \sim O_1(1). \tag 3.11
$$
We derive:
$$
O_1(1)(A+c^{'}) r^{-1} O_1(1) F_a^{[-1]} \bar{F}_n^{[1]}=O_1(1)(A+
c^{'})(r^{-1}p^{-1}Ar^{-1}+ cr^{-1} p^{-1} r^{-1})F_a^{[-1]}\bar{F}_n^{[1]},
$$
$$
\align Ar^{-1}F_a^{[-1]}\bar{F}_n^{[1]}
&=r^{-1}F_a^{[-1]}A\bar{F}_n^{[1]}+cr^{-1}F_a^{[-1]}\bar{F}_n^{[1]}\\
&+\frac{1}{at}(\partial_r F_a^{[-1]})\bar{F}_n^{[1]}\\
&=\frac{1}{at} F_a^{[-2]}(Rt)2^{-n}\bar{F}^{[2]}
+\frac{c}{at}F_a^{[-2]}\bar{F}_n^{[1]}+\frac{1}{at}(\partial_rF_n^{[-1]})
\bar{F}_a^{[1]}\\
&=O(\frac{1}{at})+O(2^{-n}\frac{R}{a}F_a^{[-2]}\bar{F}_n^{[2]}).
\endalign
$$
$$
\partial_\lambda F(\lambda=x)\equiv\frac{\partial F}{\partial
\lambda}\bigg|_{\lambda=x}.
$$
Using that $ A r^{-1} p^{-1}=O_1(1)+O_0(1)\equiv O_{0,1}(1)$, we have,
$$
\align \frac{R}{a}E_n O_1(1)F_a^{[-1]}\bar{F}_n^{[1]}
&= \frac{R}{a}E_n|p|^{-1} O_1(1)(r^{-1}|p|^{-1}A+c^{'})r^{-1}F_a^{[-1]}\bar{F}_n^{[1]}  \\
& \quad + \frac{R}{a}E_n|p|^{-1}O_{1}(1)^2 r^{-1} F_a^{[-1]}A\bar{F}_n^{[1]} \\
&= O\left(\frac{R}{a}\frac{2^n}{at}\right) + E_n\left(\frac{R}{a}
\right)^2 O_{1}(1)^2 F_a^{[-2]} \bar{F}_n ^{[2]}.\tag 3.12
\endalign
$$

 where we used that, $(L=0)$
$$
    E_n |p|^{-1} = E_n O (2^{-n}), \quad \quad r^2 \Delta = A^2 + cA + c^{'},
$$
$$
    \frac{A}{R2^{-n}t} \bar {F}_n = O(1),
$$
$$
    |p|^{-1} A r^{-1} = O(1).
$$
We now do this computation again, this time, for the term of order
$(\frac{R}{a})^2$: \newline Doing it $k$ times, we get:
$$
    \left(\frac{R}{a}\right)^k E_n O(1)^k F_a^{[-k]} \bar{F}_n^{[k]} E_n + \sum_{j=1} ^k E_n O\left(\frac{2^n}{t}
    \right) \left(\frac{R}{a}\right)^{(j-1)} \frac{1}{a} O(1)^j \tilde{F}_a^{[-j]} \bar{F}_n^{[j-1]} ,\tag 3.13
$$
with,
$$
\tilde{F}_a^{[-k]} = \frac{at}{r} F_a^{[-k+1]} + r\partial _r
\frac{at}{r} F_a ^{[-k+1]}. \tag 3.14
$$

For \quad $k \sim \delta \ln t,$ \quad and \quad $\frac{R}{a} < 1$ \quad sufficiently small,
 we have that the  $E_n F_a \bar{F}_n $ \quad term is bounded by \quad
 $O(t^{-1}2^n)$.$\blacksquare$
 \enddemo
 We can now prove the Theorem on Maximal velocity bound.
\smallskip
 {\bf Proof of Maximal Velocity Bound}
\medskip

Now,
$$
\align
&\left(\psi (t), F_a^2 (\frac{r}{t}>a) \psi(t) \right) \\
&\quad \quad=\sum_n \left(H^{-1/2} \psi(t), F_a ^2 E_n H^{1/2} \psi(t)\right) + Q \\
&\quad \quad= \sum_n \left( H^{-1/2} \psi(t), F_a ^2(F_n  (A/t) + \bar{F}_n (A/t)) E_n H^{1/2} \psi(t) \right) + Q \\
&\quad \quad= \sum_n ( H^{-1/2}\psi (t), F_a^2F_nE_n H^{1/2}\psi (t)) +
\sum_n(H^{-1/2}\psi (t), F_a^2\bar{F}_nE_n H^{1/2}\psi (t) ) + Q .\tag 3.15
\endalign
$$

%45,46,47 red
First, we use Proposition 3.1 to control the $\bar{F}_n$ term on the
RHS of (3.15)

$$
|(H^{-1/2}\psi (t), F_a^2\bar{F}_n (A/t) E_n H^{1/2} \psi (t))|
$$
$$
\leq c\| \langle x \rangle^{1/2} \psi (1) \| \left[ \|\bar{F}_n F_a^2
E_n\|  \|E_n H^{1/2} \psi(1) \|+ \| \left[ F^2_a,
\bar{F}_n(A/t)\right] E_n H^{1/2} \psi(t) \|\right], \tag 3.16
$$
where we used that
$$
\|H^{-1/2} \psi(t) \| = \| H^{-1/2} \psi(1) \| = \| H^{-1/2}\langle x
\rangle^{1/2} \langle x
\rangle^{-1/2} \psi(1) \| \leq c\| \langle x \rangle^{1/2} \psi(1) \|.
$$
By Proposition 3.1,
$$
\|\bar{F}_n(A/t) F_a^2 E_n\| \leq O(2^n/t),
$$
hence, the first term in the square bracket of (3.16) is bounded by
$$
 c2^n/t \|H^{1/2} E_n H^{1/2 + \varepsilon} H^{-1/2 -
\varepsilon} \psi (0) \|
$$
$$
\leq 0 (1) \|E_n \psi\| 2^{-n/2} + c2^{-\varepsilon n} /t \|
<x>^{1/2 + \varepsilon}\psi(0)\|.
$$
The second term in the square bracket is
$$
\| O \left(\frac{2^n}{t}\right) E_n H^{1/2} \psi(t) \|\leq
\frac{c2^n}{t} \|E_n H^{1+\varepsilon} H^{-1/2-\varepsilon}\psi (t)
\|\leq \frac{c}{t} 2^{-\varepsilon n} \|\langle x \rangle^{1/2 +
\varepsilon} \psi (1) \|.
$$
Therefore, the sum over $n$ of the $\bar{F}_n$ term in (3.15) is
bounded by

%\newline

$$
\left(o(1) + O\left(\frac{1}{t}\right)\right) \|\langle x
\rangle^{1/2+ \varepsilon} \psi (0)\| \|\langle x
\rangle^{1/2} \psi (0)\|,\text{as to }t\to \infty.
$$

$$
\align
&|\sum_n(H^{-1/2}\psi (t), F_a^2F_nE_n H^{1/2} \psi (t))| \\
&\quad \quad \quad \quad \leq \|F_a H^{-1/2} \psi(t) \| \sum_n<n>^{-1/2-\varepsilon/2} <n>^{1/2+\varepsilon/2}
 \|F_n E_n H^{1/2} \psi(t)\| \\
&\quad \quad \quad \quad \leq c\|F_a H^{-1/2} \psi(t) \|
 \left( \sum_n <n>^{1+\varepsilon} \| F_n E_n H^{1/2}
 \psi (t)\| ^2 \right) ^{1/2}\\
&\quad \quad \quad \quad \leq c\|F_a H^{-1/2} \psi(t)\|
 \left(\sum_n o(1) \| H^{1/2} E_n <x>^{1/2} \psi(1) \|^2 <n>^{1+\varepsilon }\right)^{1/2} \\
&\quad \quad \quad \quad \leq c\|F_a H^{-1/2} \psi (t)\| o(1)
\|<x>^{1/2} \psi(1) \|, \tag 3.17
\endalign
$$

$$
\align
\left|\sum_n ( H^{-1/2} \psi (t), F_a^2 \bar{F}_n E_n H^{1/2} \psi(t))\right|
&\leq \left|\sum_n (F_a H^{-1/2} \psi (t), \bar{F}_n F_a E_n H^{1/2} \psi(t))\right|\\
& + \left|\sum_n(F_a H^{-1/2} \psi (t), O(\frac{1}{at})(R2^{-n})^{-1} O(1) E_n H^{1/2} \psi(t)) \right| \\
& \leq c\left\|F_a H^{-1/2} \psi(t) \right\| \sum_n \left\|2^{+n/2} E_n
\psi(t) \right\| \frac{1}{t}\frac{R}{a}. \tag 3.18
\endalign
$$
It follows that,
$$
\align
&\left(\psi(t),F_a^2 \psi(t)\right) \leq c\left\|F_a H^{-1/2} \psi(t) \right\|
 \quad \left\| |\ln H|^{\frac{1+\varepsilon}{2}} H^{-1/2} <x>^{1/2}
 \psi_0 \right\| o(1) + Q,\\
&\|F_a H^{-1/2} \psi(t)\|\leq \| H^{-1/2}  \psi(t) \| = \| H^{-1/2}
 \psi(0)\| \leq c \| \langle x \rangle^{1/2} \psi(0) \|\tag 3.19\endalign
$$
$$
    Q \equiv \left( H^{-1/2} \psi(t), \left[H^{1/2}, F_a^2\right] \psi(t)\right).
$$
To control Q,  we need to commute fractional powers of $H$.
To this end we use that:
$$ H^\alpha P_c(H)= c_\alpha \int_0^\infty \frac{\lambda^{\alpha-1}}{\lambda+H}HP_c(H) d\lambda, $$
and estimate,
$$
\align
\lambda^{\alpha -1}\left[\frac{H}{H+\lambda},F_a\right]
&= \lambda^{\alpha -1} H \left[\frac{1}{H+\lambda},F_a \right] + [H, F_a]\frac{\lambda^{\alpha -1}}{H+\lambda} \\
&= \frac{-H}{H+\lambda} O\left(\frac{1}{t}F_a^{\prime}\right)
\frac{\lambda^{\alpha -1}}{H+\lambda} + O\left(\frac{1}{t}F_a^{\prime}\right)
\frac{\lambda^{\alpha -1}}{H+\lambda} \equiv \circledast\\
 \int_0 ^\infty d\lambda \circledast &= O(1) O\left(\frac{1}{t}F_a^{\prime}\right)O(H^{\alpha-1}). \tag 3.20
\endalign
$$
Therefore, using (3.20) with $\alpha=1/2,$we have that

$$
|Q|=|(H^{-1/2}\psi(t),O(1)O(\frac{1}{t})H^{-1/2}\psi(t))|\le\frac{c}{t}\|H^{-1/2}\psi(t)\|^2
\le\frac{c}{t}\|<x>^{1/2}\psi(0)\|^2.\tag 3.21
$$
End of Proof.

\head {Section 4. Minimal Velocity bounds}
\endhead
\medskip

We will use the analysis developed so far to obtain lower bounds on
the speed of propagation.  For this, we consider the family of
observables $ $ given by

$$
\Phi_n \equiv E_{I_n} F_n \left( \frac{A}{tR2^{-n}}\ge 0 \right)E_{I_n},
0\le R<1.\tag 4.1
$$

$$
E_{I_n} =E_{I_n}(H) = E(H\in I_n), \quad n\ge 0.
$$
Equation (0.12) applies, and we get
%3/16
$$
\partial_t\langle \psi(t), \Phi_n \psi(t)\rangle =\tag 4.2
$$
$$
=\langle \psi (t), \left\{ i [H, \Phi_n] + \frac{\partial
\Phi_n}{\partial t} \right\} \psi(t)\rangle \tag 4.2$a$
$$
$$
= \frac{1}{t} \langle E_n \psi (t) \left[ \frac{-A}{R2^{-n}}
F^\prime_n\left(\frac{A}{tR2^{-n}}\right) + |p|^{1/2} F^\prime_n(A/t)
|p|^{1/2} \frac{1}{R2^{-n}}\right. \tag 4.2$b$
$$
$$
\left.
+ 2R e|p|^{1/2} R_2 (A/t) + i [tV, F_n]\right]E_n \psi (t) \rangle
\tag 4.2$c$
$$

$$
= \frac{1}{t} \langle \psi (t), E_n \left\{ u F^\prime_n (u) |_{u =
A/Rt2^{-n}} + |p|^{1/2} F^\prime_n |p|^{1/2}\frac{1}{R 2^{-n}} \right\}
\psi (t)\rangle\tag 4.2$d$
$$
$$
+ \frac{1}{t} \langle E_n \psi(t), 2Re |p|^{1/2} R_2 (A/t) E_n \psi
(t)\rangle + \langle E_n \psi (t), i [V, F_n] E_n \psi (t)\rangle
\equiv B_{ps} + B_{re} + B_V.
\tag 4.2$e$
$$

\proclaim{Theorem 4.1}
We assume the previous notation, and the hamiltonian $H$ is as before. Then,
for all $n> N(\| | V | \|)$,
$$
\int_1^\infty \| \tilde{F}_n (A/t) E_n (H) \psi (t) \|^2 \frac{dt}{t}
< C < \infty\tag 4.3
$$
For
$$
\tilde{F}_n (u) = F(|u|< (1-\varepsilon) 2^{-n})\tag 4.4
$$

\endproclaim

\demo{Proof}
The result will follow by integration from equation (4.2) if we can
show that:

(a)
$$
\align
  &B_{ps} \geq \frac{c}{t} \langle E_n \psi (t), \tilde{F}^2_n (A/t)
E_n \psi (t) \rangle + B^1_{ps} (t), c > 0;\\
  &  \text{ for all $n$ large enough, depending only on norms of } V. \\
  & B^1_{ps}(t) \in L^1 (dt) :
\int^\infty_1 B^1_{ps} (t) dt \leq C\| E_n \psi(0) \|^2,\tag 4.5
\endalign
$$

(b)

$$
 \int^T_1 B_{re} dt \leq c \|E_n \psi(0) \|^2\tag 4.6
$$

(c)

$$
 B_V \leq \delta_0 B_{ps} + B^1_V (t) , \delta_0 \leq 1/2, \text{
and } \int^T_1 B^1_V(t) dt \leq c \|E_n \psi(0) \|^2.\tag 4.7
$$
First, we prove assertion (a).

Since $F^\prime_n (A/t)$ is a bump function localizing
$$
\frac{A}{tR2^{-n}} \sim 0,
$$
it follows that the first term of (4.2$b$) is bounded below by
$$
\frac{-A}{tR2^{-n}}  F^\prime_n \geq - \delta \tilde{F}^2_n \left(
\frac{A}{tR2^{-n}} \sim 0\right).\tag 4.8
$$
$\delta$ is a small number depending on the sharpness $\eta$ of the
function
$$
\align
&\tilde{F} (x \sim 0)^2 \equiv F^\prime (x \geq 0);\\
F\in C^\infty (\Bbb R), \quad  F_0(x\geq 0)
&= \cases \frac{1}{2\eta} + x/\eta & |x| < \eta/ 2\\
0 & x\leq -\eta/2\\
1 & x\leq \eta/2\endcases\tag 4.9
\endalign
$$
$$
F\equiv g_{\eta/10} * F_0
\tag 4.10
$$
where $g_{\eta/10}$ is an approximate $\delta$-function with support
size $\eta/10$.

The second term that contributes to (4.2$b$) is
$$
\frac{1}{t} \frac{2^n}{R} \langle \psi (t), E_n |p|^{1/2} F^\prime_n
(A/t) |p|^{1/2} E_n \psi(t) \rangle = \tag 4.11
$$
$$
\frac{1}{t} \frac{2^n}{R} \langle \psi (t), \tilde{F}_n E_n
|p|E_n\tilde{F}_n \psi(t)\rangle\tag 4.12
$$
$$
+\frac{1}{t} \frac{2^n}{R}\left\{ \langle \psi(t),
E_n\left[[E_n|p|^{1/2}, \tilde{F}_n] \tilde{F}_n |p|^{1/2} +
\tilde{F}_n E_n |p|^{1/2} [\tilde{F}_, E_n |p|^{1/2}]\right]E_n \psi
(t)\rangle\right.
$$
$$
\geq \frac{1}{t}\frac{1}{R}\langle \psi(t), \tilde{F}_n E_n
 E_n \tilde{F}_n \psi (t)\rangle
$$
$$
+ \frac{1}{t}\frac{2^n}{R}\langle \psi (t), E_n\left(C_n\tilde{F}_n
|p|^{1/2} - \tilde{F}_n E_n |p|^{1/2} C_n\right)E_n \psi (t) \rangle
\tag 4.13
$$
with
$$
C_n \equiv [E_n |p|^{1/2}, \tilde{F}_n].\tag 4.14
$$
We used in the above the following {\it Quantitative Mourre Estimate}:
$$
E_n(H)|p|E_n(H) = E_n(H)H E_n(H) -E_n(H)V E_n(H)\ge E_n(H)(2^{-n}-c<x>^{-1-\varepsilon})E_n(H)
$$$$
\ge (1-\varepsilon)E_n(H) 2^{-n}E_n(H)
$$
since $\|E_n(H)<x>^{-1-\varepsilon}\le 2^{-n(1+\varepsilon)}.$

Next, we need the following Proposition, showing that $C_n$ is higher
order correction (in t).

\proclaim{Proposition 4.2}  For $\tilde{F}_n$ defined as above, we
have

a)

$$
 [E_n(H) |p|^{1/2}, \tilde{F}_n (A/t)] = O (\frac{1}{t})
2^{n/2}  \tag 4.15
$$

b)
$$
 i[E_n (H), A] = HE^\prime_n (H)+ \tilde{E}_n (H) O (\|\langle x
\rangle^{-\sigma} \tilde{E}_n(H)\|) O(\int |\lambda \hat{E}_n
(\lambda) |)=O(1)
\tag 4.16
$$

c)

$$
[E_n(H), \tilde{F}_n(A/t)] = O(2^n/t) \tag 4.17
$$

\endproclaim

The proof of this Proposition is postponed to the end of the proof of
theorem.

For $t\leq K2^n,\text{ we have that }(\psi(t) = E_n \psi(t)
)$

$$
\align
& \partial_t\langle \psi(t), \Phi(t) \Phi(t)\rangle - \langle \psi (t),
i [H, \Phi(t)]\psi (t) \rangle\\
& = \langle \psi (t), \frac{\partial \Phi}{\partial t} \psi (t)
\rangle\\
& \sim - \langle \psi (t), \delta \tilde{F}^2_n \left(
\frac{A}{Rt2^{-n}} \sim 0\right)\psi(t) \rangle\tag 4.18
\endalign
$$

Integrating on the interval $1\leq t \leq 2^n$, we have that
%3/17
$$
\delta \int^t_1 \langle \psi(t), \tilde{F}^2_n(A/t) \psi (t) \rangle
dt
$$
$$
+ \langle \psi(T), \Phi(T) \psi(T) \rangle - \langle \psi (1), \Phi
(1) \psi(1) \rangle
$$
$$
\leq \int^T_1 |\langle \psi(t) (H E_n \Phi - \Phi E_n H) \psi (t)
\rangle | dt
$$
$$
\leq \int^{K 2^n}_1 \| E_n \psi (t) \|^2 \| H E_n \| dt \leq CK \| E_n
\psi (0) \|^2.\tag 4.19
$$
For $t\geq K 2^n$, we need to bound the second term in the expression
(4.13) by an $L^1(dt)$.

By using the proposition, part (a) it follows that
$$
(4.13)\geq \frac{1}{Rt} \langle \psi (t),\tilde{F}_n E_n \tilde{F}_n \psi (t)
\rangle
$$
$$
- 2 \frac{2^n}{Rt} \|E_n \psi (t) \|^2  \| C_n \| \| |p|^{1/2} E_n \|
  \tag 4.20
$$
$$
\geq - C \frac{2^n}{(Rt)^2} \|E_n \psi (t) \|^2 + \frac{1}{Rt} \langle
  \psi (t),  \tilde{F}_n E_n \tilde{F}_n \psi (t)
\rangle.
$$
The integral of first term on the RHS of (4.20) is bounded by
$$
\int^\infty_{K 2^n} C \frac{2^n}{(Rt)^2} \|E_n \psi (t) \|^2 dt \leq
\frac{C}{K R^2} \|E_n \psi(0) \|^2.\tag 4.21
$$

Using part (c) of Proposition (4.2), we get that
$$
\align
\tilde{F}_n E_n \tilde{F}_n &= E_n \tilde{F}^2_n + [\tilde{F}_n,
E_n]\tilde{F}_n\\
&= E_n \tilde{F}^2_n + O (2^n /t) \tilde{F}_n.\tag 4.22
\endalign
$$
Therefore
$$
\align
\frac{1}{Rt} \langle \psi(t), \tilde{F}_n E_n \tilde{F}_n\psi (t)
\rangle &= \frac{1}{Rt} \langle E_n \psi (t) ,\tilde{F}^2_n E_n \psi (t)
\rangle\\
&+ \frac{2^n}{R t^2} \langle \psi (t), O (1) \tilde{F}_n E_n\psi (t)
\rangle\tag 4.23
\endalign
$$

The $ L^1$ norm $L^1( K2^n, \infty)$ of the second term of the RHS of
equation (4.23) is bounded by $\frac{C}{RK} \|E_n \psi(0)\|^2$ as in (4.21).

This proves assertion (a).

Assertion (b) is the integrability, up to small corrections of the
$R_2(A/t)$ term, and it follows from the proof of the maximal velocity
estimate, Proposition (3.1)

Assertion (c) is control of the potential term, by an $L^1(dt)$ part
and the first term, the first term being
$$
\frac{\delta}{t} \langle \psi (t), E_n \tilde{F}_n (A/t) E_n \psi
(t)\rangle
$$

First, we rewrite the potential term.

$$
\align
&\langle \psi(t), i [H, E_n F_n E_n] \psi (t) \rangle\\
&
= \langle \psi(t) i E_n [H, F_n] E_n \psi (t) \rangle \tag 4.24
\endalign
$$

So, the potential term is
$$
\align
V_n &\equiv \langle \psi (t), i E_n [V, F_n] E_n \psi (t)\rangle\\
    &= \langle \psi (t) i E_n (\tilde{E}_n V F_n - F_n V \tilde{E}_n) E_n
\psi (t) \rangle \\
    &\text{ with } E_n \tilde{E}_n = E_n.\tag 4.25
\endalign
$$
Then, commute another $\tilde{E}_n$ through the $F_n$.

$$
\align
V_n &= \langle \psi (t), i E_n(\tilde{E}_n V\tilde{E}_n F_n +
\tilde{E}_n V [F_n, \tilde{E}_n] \\
    &- F_n \tilde{E}_n V\tilde{E}_n -
[\tilde{E}_n, F_n] V \tilde{E}_n)E_n \psi (t)\rangle\\
    &= \langle \psi (t), i E_n\left\{ [\tilde{E}_n
V\tilde{E}_n, F_n] + \tilde{E}_n V [F_n, \tilde{E}_n] \right.\\
    &- [\tilde{E}_n,
F_n] V \tilde{E}_n) E_n \psi(t) \rangle
\tag 4.26
\endalign
$$
Since by Proposition (4.2) part (c)
$$
\|[\tilde{E}_n, F_n (A/t)] \| \leq  C2^n/t\tag 4.27
$$
$$
\text{ and }
\|V\tilde{E}_n \|\leq C2^{-n (1+\varepsilon)} \text{ for } \tag 4.28
$$

$$
|V| = O (\langle x \rangle^{-\sigma}), \sigma > 1. \tag 4.29
$$
These terms decay like $2^{-\varepsilon n} t^{-1}$.

It remains to estimate $[\tilde{E}_n V \tilde{E}_n, F_n (A/t)]$.

$$
\align
& i[A, \tilde{E}_n (H) V \tilde E_n (H)] =\\
&= i[A, \tilde{E}_n (H)] V \tilde{E}_n (H) + \tilde{E}_n i [A, V]
\tilde{E}_n + \tilde{E}_n V i [A, \tilde{E}_n (H)]\\
&= \tilde{E}_n O(\langle x \rangle^{-\sigma}) \tilde{E}_n + O(1) O
(\langle x \rangle^{-\sigma}) \tilde{E}_n + \tilde{E}_n O(\langle x
\rangle^{-\sigma})O(1) \\
&= O (2^{-n(1+\varepsilon)})\tag 4.30
\endalign
$$
for $\sigma > 1$.

$\varepsilon$ can be taken to be $\frac{1}{2} -0$, provided $\sigma
\geq 3/2$.

Using (4.30), we obtain
$$
\| { F_n (A/t), \tilde{E}_n (H) V \tilde{E}_n (H) } \|=
$$
$$
 \int \hat{F}_n (\lambda) e^{i \lambda A/t} \int^\lambda_0
e^{-i s A/t}\frac{1}{t} [A, \tilde{E}_n (H) V \tilde{E}_n (H)],e^{+i s
A/t} d sd\lambda\|
$$
$$
\leq c_V\int |\hat{F}_n (\lambda) \lambda| d \lambda
2^{-n(1+\varepsilon)} / t
$$
$$
\leq c_V (2^n/Rt) 2^{-n (1+ \varepsilon)}   = c_V 2^{-\varepsilon n}/Rt.
$$

So, the potential term is higher order, and the proof is complete.
\enddemo

$\blacksquare$

{\it Proof of Proposition (4.2)}

Part (a)
$$
E_n (H) |p|^{1/2} = O (2^{-n/2}).
$$
Therefore
$$
[ E_n (H) |p|^{1/2}, \tilde{F}_n (A/t)] =
$$
$$
i\int{\hat{\tilde{F}_n}}(\lambda) e^{i\lambda A/t}\int^\lambda_0 e^{-is
A/t} \frac{1}{t} [ A, E_n (H) |p|^{1/2}] e^{-is A/t} dsd\lambda
$$
$$
[A, E_n (H) |p|^{1/2}] = [ A, E_n (H)] |p|^{1/2} + E_n (H)[A,
|p|^{1/2}]
$$
$$
= [ H E^\prime_n (H) \tilde{E}_n (H)+ O (2^n) \|\langle x
\rangle^{-\sigma}\tilde{E}_n (H)H^{1/2} \|\|H^{-1/2} |p|^{1/2}\|]\tilde E_n |p|^{1/2} + \frac{i}{2} E_n
(H)|p|^{1/2}
$$
$$
= O(2^{-n/2})+ \|\langle x\rangle^{-\sigma} \tilde{E}_n (H) \|O(2^{-n/2}) +
O(2^{-n/2}) =
$$
$$
O(2^{-n/2}),
$$
by part (b) and for $\sigma \geq 1$.

Therefore,

$$
\| [E_n (H) |p|^{1/2}, \tilde{F}_n (A/t)] \|
$$
$$
\leq \int|\lambda {\hat{\tilde{F}_n}}(\lambda)| d \lambda \frac{1}{t}
2^{-n/2} = O (2^{n/2} /t).
$$

Part (b)
$$
[E_n,A]=[E_n\tilde E_n,A]=[E_n,A]\tilde E_n+ |tolde E_n [E_n,A]
$$

$$
[E_n (H), A]\tilde{E}_n =\int\hat{E}_n (\lambda)e^{i\lambda H} \int^\lambda_0 e^{-isH}[H, A]
e^{isH} ds d\lambda \tilde{E}_n
$$
$$
= \int \hat{E}_n \lambda  e^{i\lambda H}\int^\lambda_0  e^{-isH}(-i)
[H-V-X\cdot \nabla V] \tilde{E}_n e^{isH} dsd\lambda
$$
$$
=  -iH E^\prime_n (H) \tilde{E}_n - i \int\hat{E}_n (\lambda)
e^{i\lambda H}\int^\lambda_0 e^{-isH} O (\langle x \rangle^{-\sigma})
\tilde{E}_n e^{isH} ds d\lambda
$$$$
= -i HE_n^{\prime}(H) \tilde E_n+O (\int |\lambda \hat E_n(\lambda)\|<x>^{-\sigma}\|)\tilde E_n.
$$

The last term is bounded by
$$
\int|\lambda \hat{E}_n (\lambda) |d\lambda \| \langle x
\rangle^{-\sigma} \tilde{E}_n (H) \|=O(2^n) O(2^{-\sigma n}), \sigma \le 3/2.
$$
which completes the proof of part (b).

Part (c)

$$
[E_n (H), \tilde{F}_n (A/ t)]=
$$
$$
\int{\hat{\tilde{F}_n}}(\lambda) e^{i\lambda A/t } \int^{\lambda}_0 e^{-isH/t}
[A/t, E_n (H)] e^{isA/t} ds d\lambda .
$$
Taking the norm, it is bounded by
$$
\int |\lambda {\hat{\tilde{F}_n}}(\lambda) | \frac{1}{t} \|[A, E_n (H)]
\| = O \left( \frac{2^n}{t}\right) O (1) .
$$
by part (b).

$\blacksquare$

%added 3 pagges 3/30/15

Next, we need the following estimate, similar to Proposition (3.1):
\proclaim{ Localization lemma}
$$
\| E_n (H) F_b\left( \frac{|x|}{t} < b\right) F_n \left(
\frac{A}{2^{-n} Rt} > 1 \right) E_n(H) \| =
O\left(\frac{2^n}{t}\right) \tag 4.31
$$

for $t$ large, provided $\frac{b}{R}< 1$.
\endproclaim
\demo{Proof}

Denoting momentarily $E_n (H) \equiv E_n$,
$$
F_b \left( \frac{|x|}{t} < b\right)\equiv F_b\text{ and } F_n \left(
\frac{A}{2^{-n} Rt} > 1\right) \equiv F_n,
$$
we have, using that $A= \frac{1}{2} (x\cdot p + p \cdot x) =
\sum^3_{j=1}  x_jp_j + c:$

$$
E_n F_b F_n E_n = E_n F_b x_j (F_n A^{-1}) p_j E_n
$$
$$
+c E_n F_b (F_n A^{-1}) E_n
$$
$$
+ \frac{2^n}{Rt} E_n F_b x_j [p_j,\tilde{F}_n] E_n \equiv E_n F_b x_j
(F_n a^{-1}) p_j E_n + I_1 + I_2 =
$$
$$
= \frac{b}{R} E_n \tilde{F}_b \tilde{F}_n(E_n H) H^{-1} \cdot p 2^n
$$
$$
+\frac{b}{R} E_n \tilde{F}_b\tilde{F}_n \cdot [p, E_n] 2^n + I_1 + I_2
$$
$$
\equiv J_1 + J_2 + I_1 + I_2
$$
where we define
$$
\tilde{F}_b = \frac{x}{bt} F_b = ( x_1, x_2, x_3) \frac{1}{bt} F_b
$$
$$
\tilde{F}_n = (Rt 2^{-n}) A^{-1} F_n
$$
and summation over $j$ is implied; $\cdot$ stands for scalar product
($\tilde{F}_b$ and $p$ are vectors).

We then have
$$
\| J_1\| \lesssim \frac{b}{R} \|E_n \tilde{F}_b\tilde{F}_n E_n
\|2^{-n} 2^{n} = \frac{b}{R} \| E_n \tilde{F}_b\tilde{F}_n E_n \|
$$

from $\| H^{-1} |p| \|\lesssim 1$.

$$
\| J_2 \| \leq \frac{b}{R} \|E_n \tilde{F}_b \tilde{F}_n \| 2^n \|[p,
E_n (H)]\| \leq \frac{b}{R} 2^{-n/2 +\varepsilon n} \|E_n \tilde{F}_b
\tilde{F}_n \|
$$

from $\|[p, E_n (H)] \| < 2^{-\left(\frac{3}{2} -\varepsilon\right)
n}$

$$
 I_1 = c\frac{2^n}{Rt} E_n F_b \tilde{F}_n E_n
$$
$$
I_2 = \frac{b}{R} E_n \tilde{F}_b \left(\frac{2^n}{Rt}\right) \cdot
2^n O \left( \tilde{F}_n^\prime\right) pE_n
$$
$$
\approx \frac{b}{R} \frac{2^n}{Rt}E_n \tilde{F}_b O \left(
\tilde{F}_n^\prime\right) E_n
$$

So, $I_1$ and $I_2$ are higher order, of the form
$$
\frac{2^n}{Rt} E_n O (1) E_n
$$
$$
J_2 \lesssim E_n \tilde{F}_b \tilde{F}_n O \left( \frac{b}{R} 2^{-
\frac{n}{2} + \varepsilon n}\right)
$$

So, $J_1$ is the leading term.

$$
| \tilde{F}_b| = | \frac{x}{bt} F \left( \frac{|x|}{t} < b \right)
  |\leq F_b \left( \frac{|x|}{t} < b \right)
$$
$$
\tilde{F}_n = A^{-1} F_n (A/t) \left(tR2^{-n}\right) \leq F_n
$$
So
$$
\|J_1\| \leq \frac{b}{R} \|E_n F_b \frac{\tilde{F}_b}{F_b} F_n
\frac{\tilde{F}_n}{F_n} E_n \|
$$

$$
\leq \frac{b}{R} \| E_n F_b F_n E_n\|
$$
$$
+ \frac{b}{R}\|[E_n, \tilde{F}_b/F_b] F_b \tilde{F}_n E_n \|
$$

$$
+ \frac{b}{R}\| E_n F_b F_n[ \tilde{F}_n / F_n, E_n] \|
$$

$$
G_n\equiv\frac{\tilde{F}_n}{F_n} = F (A > Rt2^{-n}) / (A/(Rt2^{-n}))
$$
$$ G_b\equiv \frac{\tilde F_b}{F_b}$$
and the proof follows since $[F_b,E_n]=O(2^n/t)=[G_b,E_n],$
and

$$
[F_b, E_n] \text{ and } [F_n, E_n] = O (2^n/t).
$$
\enddemo
$\blacksquare$

 The above analysis is now used to prove minimal velocity bounds, following similar arguments leading to the maximal velocity bounds:
$$
 \langle \psi (t),F_b(\frac{|x|}{t}<b)\psi(t)\rangle=\langle H^{-1/2} \psi (t),F_b(\frac{|x|}{t}<b)H^{1/2}\psi(t)\rangle+Q(t).
 \tag 4.32
 $$
 $Q(t)$ comes from commuting $H^{1/2}$ through $F_b$.
 As shown before, this term goes to zero as time tends to infinity, as in equation (4.21).
 So, we only need to estimate the first term of equation (4.32).

 $$
 \aligned
 \sum_n \langle H^{-1/2} \psi (t),F_b(\frac{|x|}{t}<b)E_n^2 H^{1/2}\psi(t)\rangle
  &=\sum_n [ \langle H^{-1/2} \psi (t),E_nF_b(\frac{|x|}{t}<b)E_n H^{1/2}\psi(t)\rangle \\
&+\langle H^{-1/2} \psi (t),O(2^n/t)E_n H^{1/2}\psi(t)\rangle ]
\endaligned
\tag 4.33
$$

As before, the second term of  (4.33) is controlled by
$$
\frac{c}{t}\|<x>^{1/2}\psi(0)\|^2.
$$

So, it remains to control the first part of (4.33).

Proceeding as before, we write in this first term
$$
E_nF_b(\frac{|x|}{t})E_n =E_nF_b(\frac{|x|}{t}<b)(F_n(\frac{A}{2^{-n}Rt})+\bar F_n(\frac{A}{2^{-n}Rt}))E_n .
$$

We have shown in the localization lemma that
$$
E_nF_bF_nE_n=O(2^n/t).
$$
The propagation estimates of section 4 imply:
$$
\|E_nF_bF_nE_nH^{1/2}\psi(t)\|=o(1) 2^{-n}\|<x>^{1/2}\psi(0)\|_2.
$$
$$
\| H^{-1/2}\psi(t)\|= \| H^{-1/2}\psi(0)\|\lesssim \|<x>^{1/2}\psi(0)\|_2.
$$

Putting it all together,this term also vanishes as time goes to infinity.
So, we finally have:

\proclaim{Theorem-Minimal Velocity Bound}
For $H$ as before, we have for $b<1$:
$$
\langle \psi (t),F_b(\frac{|x|}{t}<b)\psi(t)\rangle=o(1)\|<x>^{1/2}\psi(0)\|_2^2
$$
$$
\int_1^{\infty} \langle \psi (t),F_b(\frac{|x|}{t}<b)\psi(t)\rangle \frac{dt}{t} \leq c\|<x>^{1/2}\psi(0)\|_2^2.
$$
\endproclaim

\newpage

{\bf REFERENCES}
\medskip
[BFS]~  V. Bach, J. Fr\"ohlich, and IM Sigal.: Commun. Math. Phys.,
{\bf 207} :249-290, 1999

\medskip

[BFSS]~  V. Bach, J. Fr\"ohlich, I. M. Sigal, A. Soffer, {\it
Positive Commutators and Spectrum of Pauli-Fierz Hamiltonians of
Atoms and Molecules}, Comm. Math. Phys., {\bf 207}, 1999, 557-587.
\medskip
[FGS]~ J. Fr\"ohlich, M. Griesemer, I. M. Sigal {\it Spectral
renormalization and Local Decay...}, (2009), arxiv0904.1014v1
\medskip

[Ger] C. Gerard, JS Møller, MG Rasmussen  "Asymptotic completeness in quantum field theory: translation invariant Nelson type models restricted to the vacuum and one-particle sectors",  Letters in Mathematical Physics, 95.2 (2011): 109-134.

\medskip
 [HSS]~ W. Hunziker, I. M. Sigal, A. Soffer, {\it Minimal
Velocity Bounds}, Comm. PDE, {\bf 24}, (1999), No. 11/12, 2279-2295.
\medskip

[Sig-Sof]~ I.M. Sigal and A. Soffer, {\it Local Decay and Propagation
Estimates for Time Dependent and Time Independent Hamiltonians}, Preprint, Princeton 1988,
(ftp://www.math.rutgers.edu/pub/soffer).

\medskip
[GJY]~ Galtbayar, A., Arne Jensen, and Kenji Yajima. "The Nelson model with less than two photons." Annales Henri Poincare. Vol. 4. No. 2. Birkhäuser Verlag, 2003.

\medskip
[GZ]~ Marcel Griesemer, Heribert Zenk,  "On the Atomic Photoeffect in Non-relativistic QED"
Communications in Mathematical Physics
 Volume 300, Issue 3, pp 615-639
 2010

\medskip
[BFSig]~ Bony, Jean-François, Jérémy Faupin, and Israel Michael Sigal. "Maximal velocity of photons in non-relativistic QED." Advances in Mathematics 231.5 (2012): 3054-3078.

\medskip
[FSig1]~ Faupin, Jeremy, and Israel Michael Sigal. "On quantum Huygens principle and Rayleigh scattering."  arXiv:1202.6151 (2012).

\medskip
[FSig2]~ Jeremy Faupin, Israel Michael Sigal
"On Rayleigh Scattering in Non-Relativistic Quantum Electrodynamics"

Communications in Mathematical Physics
June 2014, Volume 328, Issue 3, pp 1199-1254

\medskip
[Kit]~ H. Kitada
"Scattering Theory for the Fractional Power of Negative Laplacian " Jour. Abstr. Differ. Equ. Appl.
Volume 1, Number 1 (2010), 1 - 26

Date: 23 Feb 2014

\enddocument

\end